\documentclass[aps,pre,amsmath,amssymb,reprint]{revtex4-1}
\usepackage[utf8]{inputenc}
\usepackage{mfirstuc}
\usepackage{graphicx}
\usepackage{subfigure}
\usepackage{float}
\usepackage{gensymb}
\usepackage{amsmath}
\usepackage{color}
\usepackage{multirow}
\usepackage{comment}
\usepackage[colorlinks=true, linkcolor=blue, citecolor=blue, urlcolor=blue]{hyperref}

\begin{document}
\title{Activity-Induced Structural and Dynamical Heterogeneity in Colloidal Gels}

\author{Santu Nath}
\thanks{These authors contributed equally to this work.}
\author{Sudipta Mandal}
\thanks{These authors contributed equally to this work.}
\author{Smarajit Karmakar}
\email{smarajit@tifrh.res.in}
\affiliation{Tata Institute of Fundamental Research, 36/P, Gopanpally Village, Serilingampally Mandal, Ranga Reddy District, Hyderabad 500046, Telangana, India}
\date{\today}

\begin{abstract} 
Active gels are intrinsically non-equilibrium materials whose structural and dynamical heterogeneity emerges from the interplay of attractive interactions, thermal fluctuations, and self-propulsion. We investigate a model active gel, simulated using large-scale Brownian dynamics simulations, to explore the effects of temperature $T$, active force strength $f_0$, persistence timescale $\tau_p$, and active particle concentration $\rho_a$. We characterize the structure of the evolving gel network through morphological analysis, while dynamical changes are quantified by the relaxation timescale and dynamical heterogeneity (DH). We show that an optimal choice of activity parameters yields a percolating gel network with faster gelation and larger voids and clusters, enhancing mechanical stability. Conversely, the structural relaxation timescale and DH decrease with increasing activity, indicating enhanced particle rearrangements and reduced spatial correlations of mobile and immobile regions. An effective temperature ($T_\text{eff}$) framework for colloidal gels maps structural quantities and relaxation timescales in an active gel onto those of a passive gel. More interestingly, the growth of DH with increasing swim force can also be described by the same effective-temperature framework, in stark contrast to observations in active glasses. Our findings show that, in colloidal gels, temperature and active force compete, giving rise to similar structural ordering and, consequently, similar DH at the same $T_\text{eff}$. These findings highlight the crucial role of structural ordering in governing DH in active colloidal gels, as compared to active glasses.
\end{abstract}
\maketitle

\section{Introduction}
Colloidal particles with strong attractive interactions can form kinetically arrested~\cite{Zaccarelli_Jour-of-Phys_2007, Zaccarelli_Nature_2008, Lu_Annual-Rev_2013, Patrick_Jour-of-Phys_2021, Weitz_PRL_2001} and non-ergodic disordered states of matter~\cite{Tanaka_PRE_2004, Ren_PRL_1993}, characterized by heterogeneous network structures with highly porous spaces~\cite{Weitz_PRL_2006}, commonly referred to as \textit{gels}. These states of matter are ubiquitous in soft condensed matter physics~\cite{Bethier_PRM_2017} and exhibits out-of-equilibrium or aging dynamics~\cite{Ranjhini_Soild-State-Comm_2006, Tanaka_Nat-Phys_2024}, a process commonly known as $``$\textit{gelation}$"$~\cite{Tanaka_PRE_2004}. Gels are found in everyday materials such as toothpaste and yogurt, and they have wide-ranging industrial applications in paints~\cite{Dan_Advanced-Materials_2019}, medicine~\cite{Luca_Gels_2020}, cosmetics~\cite{Claudia_Gels_2025}, batteries~\cite{Larsen_Rheologica-Acta_2024}, and biomedical engineering~\cite{Rose_Nature_2014}. Beyond these conventional applications, incorporating self-propelled agents~\cite{Cates_Arxiv_2025} into a percolating gel network offers a route to self-driven, smart, and disordered materials. Doping active particles into a colloid-polymer mixture~\cite{omar_acs-nano_2018, chandana_prl_2024, brady_pnas_2024, mallory_jcp_2020, pietro_pnas_2018, olivier_prl_2023, Solomon_Soft-Matter_2019} introduces energy consumption and persistent structural rearrangements, thereby modifying the structure and dynamics of the percolating gel network. Active gels are fundamental in several biological systems such as actomyosin networks in cells~\cite{backouche_phys-bio_2006}, the cytoskeleton of the eukaryotic cell~\cite{swartz_soft-matter_2021}, mucus and cilia-driven biological gels~\cite{lynda_pnas_2017}, aggregation of the soil bacterium Myxococcus xanthus~\cite{copenhagen_prx-life_2025}, etc.

\vskip+0.04in
Several studies on the doping of active agents into colloidal gels, including simulations~\cite{omar_acs-nano_2018, chandana_prl_2024, brady_pnas_2024, mallory_jcp_2020} and experiments~\cite{pietro_pnas_2018, olivier_prl_2023, Solomon_Soft-Matter_2019}, have predominantly focused on the structural evolution and mechanical responses of the medium. However, studies on active glasses highlight different aspects, such as dramatic growth of dynamical heterogeneity (DH)~\cite{Paul_PNAS_2023}, tunable mechanical responses~\cite{Sharma_Nat-Phys_2025, Priya_Nat-Comm_2026}, and enhanced long-wavelength fluctuations~\cite{Dey_Nat-Comm_2025}. Glasses exhibit pronounced dynamical heterogeneity~\cite{Berthier_Physics_2011} without discernible structural changes, whereas gels show a striking coexistence of structural and dynamical heterogeneity, arising from their intrinsically heterogeneous, hierarchical, system-spanning network structures. Several experimental~\cite{Solomon_PRE_2008} and simulation studies~\cite{Cates_Jour-Phys-Chem-B_2005, Gado_PRE_2008, Michael_JCP_2004} have established the existence of DH in colloidal gels.

\vskip+0.04in
Dynamical heterogeneity in glasses arises from heterogeneous caging effects~\cite{Kob_PRE_1997}, leading to the concurrent occurrence of fast and slow moving particles. Conversely, DH in colloidal gels arises from the presence of distinct mobile (fast) and immobile (slow) particle populations, resulting from exchange between bound and free particles within heterogeneous structures~\cite{Cates_JCP_2004}. The exchange between these bimodal populations of fast and slow particles is slow. Fast particles experience a rapidly relaxing local environment and have fewer neighbors, whereas slow particles form persistent bonds with themselves and have a larger number of neighboring particles~\cite{Cates_JCP_2004, Cates_Jour-Phys-Chem-B_2005}. Thus, the mobile particles move through the interconnected network structures formed by the immobile particles, resulting in spatially heterogeneous dynamics. In turn, some regions exhibit higher particle accumulation, while others have lower particle concentrations, resulting in pronounced structural heterogeneity. Ref.~\cite{jamali_soft-matter_2024} shows that variations in interparticle interactions can induce structural heterogeneity in the colloidal gel network.

\vskip+0.04in
In active glasses, doping with self-propelled agents has been shown to enhance DH and show no discernible changes in the structures with increasing swim forces, while preserving a similar relaxation timescale~\cite{Paul_PNAS_2023, Sussman_PRE_2025}. However, despite the intrinsically ramified structures of colloidal gels, the influence of activity on their DH remains largely unexplored. This work shows how active motile particles play a pivotal role in modifying the structural and dynamical heterogeneity of colloidal gels and examines the interplay between these two forms of heterogeneity as a function of swim force, which have similar structural relaxation timescales.

\begin{figure*}[!htpb]
    \centering
    \includegraphics[page=1, trim=0cm 3.3cm 0cm 0cm, clip, width=1.00\textwidth]{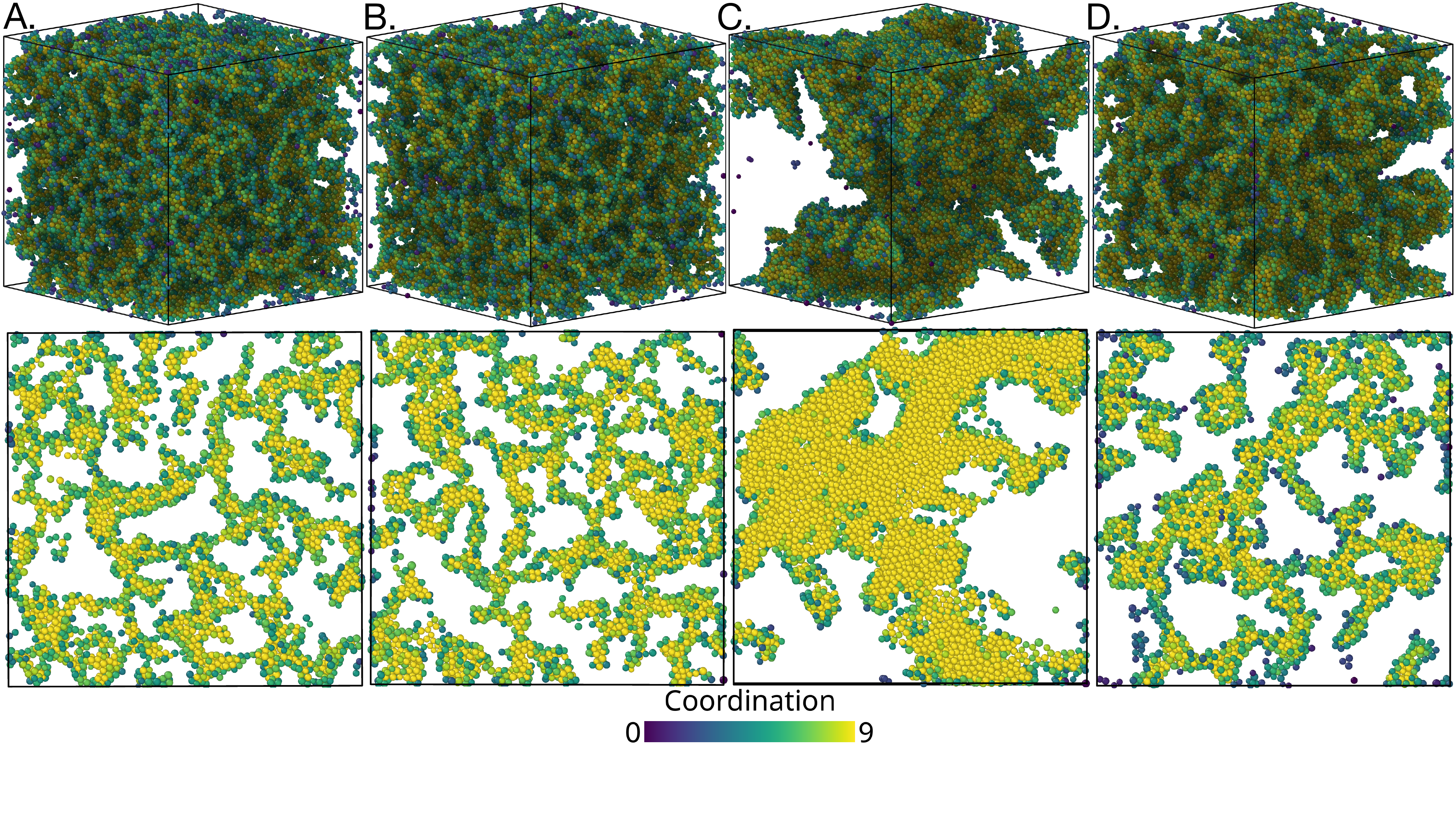}
    \caption{\textbf{Structural evolution:} Representative snapshots illustrating the percolating gel network structure under different activity conditions. (A) The top panel shows a purely passive gel system at $t \sim 2 \times 10^5$, simulated at $T = 0.10$, while the bottom panel shows its cross-section view, revealing a nested percolating network. The color scale represents the coordination number $z$ between $0$ (blue) and $9$ (yellow). (B) Activity doping transforms the gel into an active system, simulated at $T = 0.10$, with the following activity parameters $f_0 = 7.00$, $\rho_a = 0.10$, and $\tau_p = 10^0$. The cross-sectional view shows that activity leads to a slightly aggregated network structure compared to the purely passive gel. (C) Shows an active gel system simulated with the same $f_0$ and $\rho_a$ as in (B), while $\tau_p$ is varied to $10^6$. Increasing $\tau_p$ results in a significantly larger and wider percolating gel network. (D) Making all particles active in the network, with $f_0 = 7.00$ and $\tau_p = 10^0$, results in a slightly more aggregated network structure compared to (B).}
    \label{fig_structure}
\end{figure*}

\vskip+0.04in
Embedding a small fraction of self-propelled agents with additional internal energy into colloidal gels can profoundly reshape the dynamical free-energy landscapes~\cite{omar_acs-nano_2018} via a thermally activated process. A natural question is arising how these thermally activated non-equilibrium states of matter is related to their passive counterparts through an appropriate thermodynamic description. The concept of an ``effective temperature" $T_\text{eff}$~\cite{Marjolein_Soft-Matter_2016, Cugliandolo_PRE_2008, Ludovic_EuroPhys-Letter_2015} has emerged as a powerful tool for characterizing non-equilibrium active systems, and it is well established in active glasses~\cite{Paul_PNAS_2023, Saroj_PNAS_2018, Saroj_Soft-Matter_2017}. However, whether and how this thermodynamic description in terms of $T_\text{eff}$ can be established for active colloidal gels remains largely unexplored. Herein, we develop an effective temperature framework for active gels that establishes a correspondence between structural quantities, including bond number, void and cluster size, and the structural relaxation timescale with a passive aged gel.

\section{Results and Discussion}
We begin by investigating the influence of activity on the structural and dynamical properties of an active gel. We first examine the evolution of key structural quantities, such as the number of bonds, gelation time, void size, and largest cluster size, as a function of $f_0$ for different choices of $T$, $\tau_p$, and $\rho_a$. We then analyze the age-dependent growth of the relaxation timescale and dynamical heterogeneity as a function of activity. Finally, we discuss the effective temperature framework for an active gel system and the relationship between structural and dynamical heterogeneity at the same $T_\text{eff}$.

\vskip+0.04in
\noindent{\bf \large Activity-driven structural evolution:}
The structure of an active gel emerges from the competing mechanisms between self-propulsion forces and thermal fluctuations, in particular the competition between the kinetics of gelation timescale and persistent active motion~\cite{chandana_prl_2024, olivier_prl_2023}, leading to distinct structural reconfigurations. To understand how activity influences structural evolution, we first examine representative network configurations across different activity conditions. Figure~\ref{fig_structure} presents configurations at time $t \sim 2 \times 10^5$, for different choices of activity parameters. The top panel shows the corresponding three-dimensional gel structures, while the bottom panel represents the cross-sectional view of the internal gel network. Figure~\ref{fig_structure}(A) shows the reference passive gel, simulated at $T = 0.10$, where attractive interactions and thermal fluctuations lead to the formation of a connected gel network with a relatively heterogeneous morphology. In Fig.~\ref{fig_structure}(B), we next introduce activity by setting $f_0 = 7.00$, with $T = 0.10$, $\rho_a = 0.10$ and $\tau_p = 10^0$. Compared with the passive gel structure, the active system exhibits a moderately more heterogeneous morphology, characterized by the emergence of slightly larger dense aggregates and redistribution of the void spaces.

\vskip+0.04in
To isolate the role of persistence time, Fig.~\ref{fig_structure}(C) shows the gel obtained for a much larger persistence time, $\tau_p = 10^6$, while keeping $T = 0.10$, $f_0 = 7.00$ and $\rho_a = 0.10$ unchanged as in Fig.~\ref{fig_structure}(B). In contrast to the configuration at the smaller persistence time $\tau_p = 10^0$, the gel network exhibits a more compact, consolidated morphology with more pronounced large-scale aggregates. Alternatively, to investigate the role of active particle concentration, we make all colloidal particles active in Fig.~\ref{fig_structure}(D), while keeping $T = 0.10$, $f_0 = 7.00$, and $\tau_p = 10^0$ unchanged, as in Fig.~\ref{fig_structure}(B). The cross-sectional view in Fig.~\ref{fig_structure}(D) shows more porous spaces than in Fig.~\ref{fig_structure}(B). This structural evolution across different sets of activity parameters indicates that the percolating network in colloidal gels is strongly influenced by changes in $\tau_p$ and $\rho_a$. This implies that an optimal combination of activity parameters exists that accelerates gel formation and promotes the development of large voids and clusters spanning the entire simulation box.

\begin{figure*}
    \centering
    \includegraphics[page=1, trim=0cm 13cm 0cm 0cm, clip, width=1.00\textwidth]{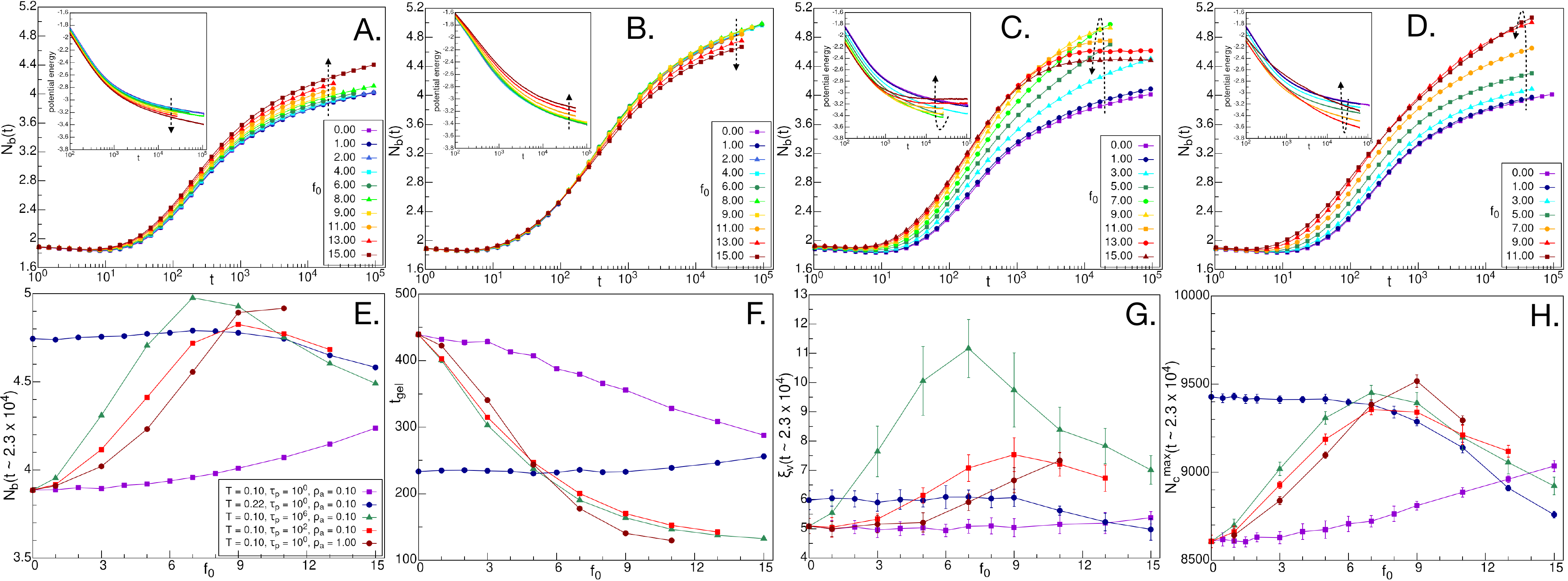}
    \caption{\textbf{Morphology of active gels:} Time evolution of average number of bonds per particle $N_b(t)$ for different $f_0$ at (A) $T = 0.10$ and (B) $T = 0.22$, with $\tau_p = 10^0$ and $\rho_a = 0.10$ held fixed. The inset shows the time evolution of the corresponding potential energy, using the same colors as in the main panel. An increase in $N_b(t)$ and a decrease in potential energy (see black dashed arrows) with increasing $f_0$ promote the formation of a more stable and well-connected gel network at $T = 0.10$. However, for $T = 0.22$ shows qualitatively opposite results. (C) Shows time evolution of $N_b(t)$ for different $f_0$ at $T = 0.10$, $\tau_p = 10^{6}$, $\rho_a = 0.10$. At the long-time limit, $N_b(t)$ initially increases and then decreases with $f_0$, indicating a non-monotonic response of the gel network. The corresponding potential energy shows a complementary trend, initially decreasing and then increasing as $f_0$ increases. (D) Time evolution of $N_b(t)$ for different $f_0$ at $T = 0.10$, $\tau_p = 10^{0}$, and $\rho_a = 1.00$, showing a similar behavior to (C). (E) compares $N_b(t)$ as a function of $f_0$ at $t \sim 2.3 \times 10^4$ for different choices of activity parameters corresponding to (A)\textendash(D). (F) shows the variation of gelation time $t_\text{gel}$ as a function of $f_0$ for the same set of activity parameters in (E). (G) and (H) show the variation of the largest void length $\xi_v(t)$ and number of isostatic particles in the largest cluster $N_c^\text{max}(t)$, respectively, with $f_0$ at $t \sim 2.3 \times 10^4$, for the same set of activity parameters as in (E).}
    \label{fig_morphology}
\end{figure*}

\vskip+0.04in
To study the structural heterogeneity~\cite{jamali_soft-matter_2024, xiaoming_soft-matter_2021} in a percolating gel network, we characterize it through different morphological analyses. We first examine the average number of bonds per particle, $N_b(t)$~\cite{Zaccone_PRE_2020, Zaccone_Nat-Comm_2020, himangsu_pre_2025}, which relates to the coordination number $z$ as $N_b(t) = z/2$, since each bond is shared by two particles. Figure~\ref{fig_morphology}(A) and \ref{fig_morphology}(B) shows the temporal evolution of $N_b(t)$, in variation with $f_0$, simulated at temperatures $T = 0.10$ and $T = 0.22$, respectively, with $\tau_p = 10^0$ and $\rho_a = 0.10$ are held fixed. At the long-time limit in Fig.~\ref{fig_morphology}(A), $N_b(t)$ increases systematically with $f_0$ (shown by the dashed black arrow), indicating enhanced network connectivity and more effective bond formation with increasing activity. In contrast, at the higher temperature $T = 0.22$, as shown in Fig.~\ref{fig_morphology}(B), $N_b(t)$ decreases at sufficiently large $f_0$, suggesting that strong activity increasingly disrupts the attractive network through enhanced bond breaking. Insets of Fig.~\ref{fig_morphology}(A) and \ref{fig_morphology}(B) show the corresponding temporal evolution of the potential energy. At $T = 0.10$, in the long-time limit, the potential energy decreases with increasing $f_0$, indicating more effective annealing and the formation of a more energetically stable, strongly connected network. However, at $T = 0.22$, the potential energy shows the opposite trend, increasing with higher activity, indicating greater disruption of the attractive network and reduced structural stability. This suggests that the balance between thermal fluctuations and self-propulsion forces controls the stability of the colloidal gel network. We find that simulations at $T = 0.10$ with largely varying activity parameters allow us to systematically explore their effects on percolating network formation.

\vskip+0.04in
Figure~\ref{fig_morphology}(C) shows the temporal evolution of $N_b(t)$ for different values of $f_0$ at a large persistence time $\tau_p = 10^6$, with $T = 0.10$ and $\rho_a = 0.10$. Consistent with the more compact morphology observed in Fig.~\ref{fig_structure}(C), in the long-time limit $N_b(t)$ increases with active force up to $f_0 \leq 7.00$, indicating enhanced network connectivity. Beyond this active force strength, $N_b(t)$ decreases, indicating increased network destabilization due to enhanced bond breaking. The corresponding potential energy evolution in the inset shows the same non-monotonic trend, suggesting the same critical value of $f_0^c = 7.00$ for the activity parameters $\tau_p = 10^6$ and $\rho_a = 0.10$ at $T = 0.10$. A similar behavior is observed for the fully active particle system ($\rho_a = 1.00$) with $T = 0.10$ and $\tau_p = 10^0$ in Fig.~\ref{fig_morphology}(D), demonstrating that excessive activity can destabilize the gel network beyond the strength of the active force of $f_0^c = 9.00$. The inset of Fig.~\ref{fig_morphology}(D) shows the temporal evolution of the potential energy for different values of $f_0$, revealing the critical active force $f_0^c = 9.00$, beyond which the potential energy increases, hindering annealing and destabilizing the gel network.

\vskip+0.04in
To obtain a systematic comparison across all activity parameters, Fig.~\ref{fig_morphology}(E) shows $N_b(t)$ as a function of $f_0$ at time $t \sim 2.3\times10^{4}$, chosen to be well beyond the gelation time $t_\text{gel}$. We observe that with increasing $f_0$, $N_b(t)$ increasing for $T = 0.10$ (violet square) and decreasing for $T = 0.22$ (blue circle) with fixed $\tau_p = 10^0$ and $\rho_a = 0.10$. Our findings also suggest that, at $T = 0.10$ and $\rho_a = 0.10$, $N_b(t)$ reaches its maximum at a critical active force $f_0^c$ that depends on the persistence time $\tau_p$. For $\tau_p = 10^6$ (green triangle), we find $f_0^c = 7.00$, while for $\tau_p = 10^2$ (red square), $f_0^c = 9.00$. In contrast, for $\tau_p = 10^0$ (violet square), the maximum occurs beyond the investigated range, or $f_0^c > 15.00$. These results indicate that increasing the $\tau_p$ shifts the $f_0^c$ toward lower values. Similarly, upon increasing $\rho_a$ to $1.00$ (brown circle), $f_0^c$ shifts to $9.00$, compared with the $\rho_a = 0.10$ (violet square) case. This indicates that increasing $\rho_a$ shifts the $f_0^c$ toward lower values.

\vskip+0.04in
We next quantify the kinetics of gel formation by calculating the gelation time $t_\text{gel}$, which provides a characteristic timescale for the development of the connected network structure. The gelation point in the experiment is characterised by oscillatory rheology~\cite{Ness_Soft-Matter_2023}, which probes the liquid-to-solid transition through measurements of the storage and loss moduli. According to the Maxwell criterion for mechanical stability~\cite{Maxwell_Magazine_1864}, we define the gelation time as the time required for the average coordination number $z$ to reach the prescribed threshold value $z^{*} = 6$, corresponding to the formation of a percolating network of isostatic particles~\cite{Van_Jour-of-Phys_2009, Zaccone_PRB_2011, Santos_PRE_2020}. This method has been validated through both experimental~\cite{Tsurusawa_Sci-Advances_2019} and simulation studies~\cite{Ness_Soft-Matter_2023, Jiang_Nat-Comm_2023}. Figure~\ref{fig_morphology}(F) shows the dependence of $t_\text{gel}$ on the active force $f_0$ for different temperatures with $\tau_p = 10^0$ and $\rho_a = 0.10$ held fixed. For $T = 0.10$, $t_\text{gel}$ decreases systematically with increasing $f_0$, indicating that stronger active forces accelerate the formation of the connected gel network. In contrast, at $T = 0.22$, $t_\text{gel}$ remains nearly insensitive to $f_0$ with low activity strength. However, beyond a certain $f_0$, $t_\text{gel}$ increases with increasing $f_0$. For sufficiently high temperature and large active forces, enhanced particle motion promotes bond breaking and network restructuring, thereby delaying gel formation. Increasing the persistence time to $\tau_p = 10^2$ and $10^6$ leads to a pronounced reduction in $t_\text{gel}$, as compared with $\tau_p = 10^0$, with $T = 0.10$ and $\rho_a = 0.10$ fixed, indicating that more persistent active motion promotes faster percolating network formation. A similar trend is observed upon increasing $\rho_a$ to $1.00$, while keeping $T = 0.10$ and $\tau_p = 10^0$ fixed, indicating that a larger fraction of active particles leads to a shorter $t_\text{gel}$. Figure~\ref{fig_morphology}(F) shows that beyond the critical active force $f_0^c$, $t_\text{gel}$ begins to saturate, suggesting that increasing activity accelerates gel formation within the stable-gel regime, whereas sufficiently strong activity ultimately destabilizes the gel network.

\begin{figure*}
    \centering
    \includegraphics[page=1, trim=0cm 4.5cm 0cm 0cm, clip, width=1.00\textwidth]{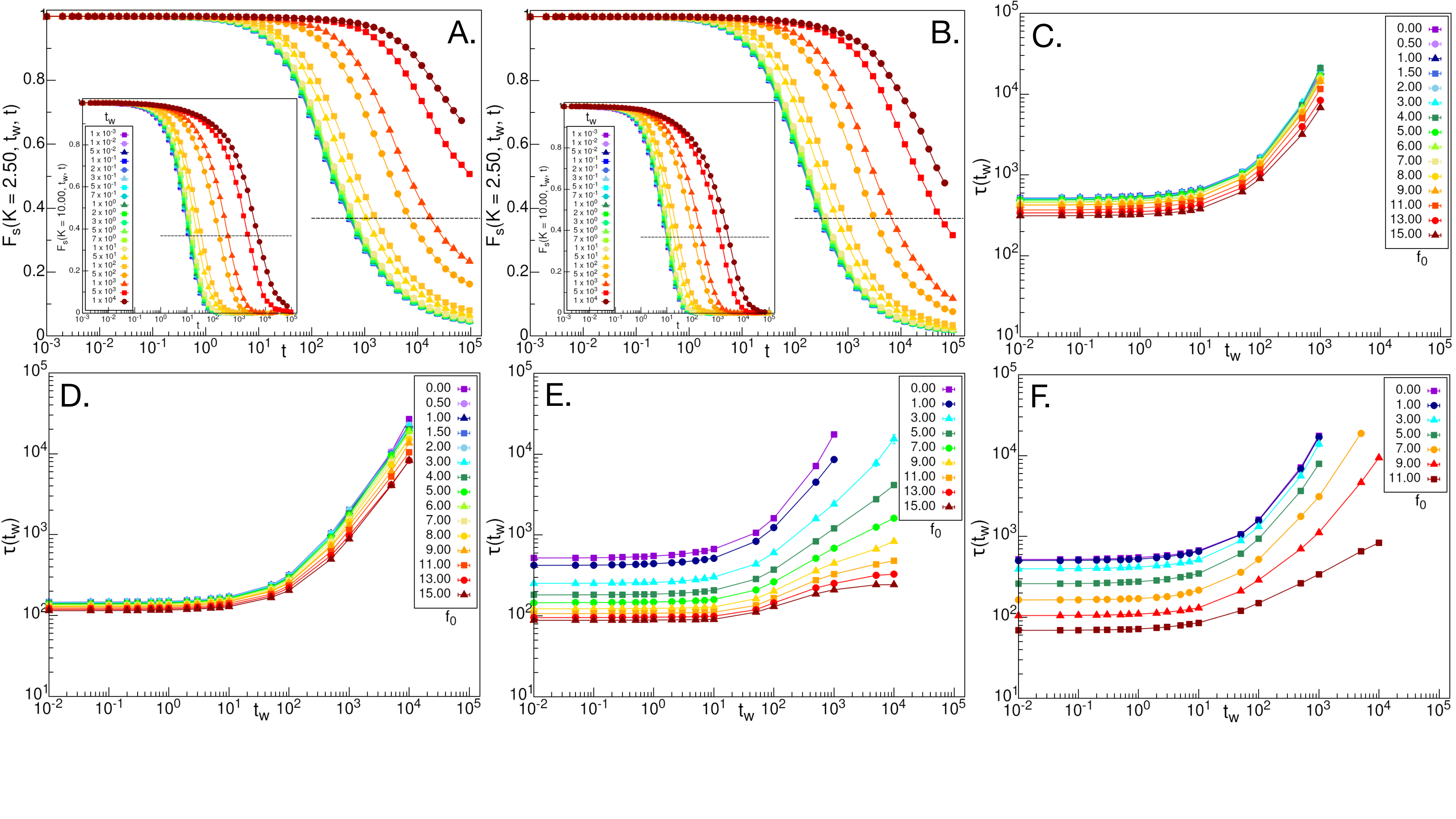}
    \caption{\textbf{Relaxation dynamics in aging:} The waiting time dependence of $F_s(K,t_w,t)$ is shown for different self-propulsion forces\textemdash (A) $f_0 = 0.00$ and (B) $f_0 = 15.00$, while $T = 0.10$, $\tau_p = 10^0$, and $\rho_a = 0.10$ are held fixed. The magnitude of the wave vector is chosen as $|\vec{K}| = 2.50$. The inset shows the temporal evolution of $F_s(K,t_w,t)$ for the same waiting times as in the main panel at a larger wave vector $|\vec{K}| = 10.00$. At a given waiting time $t_w$, increasing $f_0$ causes $F_s(K,t_w,t)$ to relax faster. The horizontal $1/e$ line in panels (A) and (B) is used to estimate the structural relaxation time $\tau(t_w)$. The waiting time variation of $\tau(t_w)$ for different choices of $f_0$ at (C) $T = 0.10$ and (D) $T = 0.22$ with $\tau_p = 10^0$ and $\rho_a = 0.10$ is shown. For a given $f_0$, increasing the temperature decreases the relaxation timescale during aging. (E) shows the variation of $\tau(t_w)$ during aging for different $f_0$ values is shown at a large persistence time $\tau_p=10^6$, in comparison with (C). For large $f_0$ values, the extracted $\tau(t_w)$ leaves the aging regime and attains a steady-state regime at the long-time limit. (F) shows $\tau(t_w)$ in variation with different $f_0$, where all particles are active, simulated at $T = 0.10$, and $\tau_p = 10^0$. Making all particles are active $\rho_a = 1.00$, leads to faster relaxation with increasing $f_0$ at a given $t_w$, compared with panel (C).}
    \label{fig_relaxation}
\end{figure*}

\vskip+0.04in
To further characterize the structural evolution and mechanical stability of the gel network, we analyze the connectivity of locally isostatic particles by evaluating the length scale associated with the largest void $\xi_v(t)$ ~\cite{Jiang_Nat-Comm_2023}, and the number of isostatic particles in the largest cluster $N_c^\text{max}(t)$~\cite{Zaccone_Nat-Comm_2020, Zaccone_PRE_2020}. In Figs.~\ref{fig_morphology}(G) and \ref{fig_morphology}(H), we show the evolution of $\xi_v(t)$ and $N_c^\text{max}(t)$, respectively, as a function of the strength of active forces at $t \sim 2.3 \times 10^4$ for different activity parameters. Following Refs.~\cite{Nick_Soft-Matter_2015, Jiang_Nat-Comm_2023} we define the void volume $V_\text{void}$ as the volume of a sphere centered at a grid point, with its radius determined by the distance from the grid point to the nearest particle surface. The corresponding void length $\xi_v(t)$ is defined as the linear dimension of the largest void, which is given by $\xi_v(t) = \sqrt[3]{V_\text{void}}$. The experimental work in Ref.~\cite{Tsurusawa_Sci-Advances_2019} suggests that the emergence of an isotropic percolating network of locally isostatic particles results in a more mechanically stable structure. Accordingly, we identify locally isostatic particles as those with $z \ge 6$ and remove particles with $z < 6$. Next, we identify clusters of isostatic particles and, as a proxy, estimate the number of isostatic particles in the largest cluster, $N_c^\text{max}(t)$, to quantify the mechanical stability of the isostatic percolating network~\cite{Ness_Soft-Matter_2023, Caporusso_Soft-Matter_2024}. Compared with the results at $T = 0.10$ and $T = 0.22$ with $\tau_p = 10^0$ and $\rho_a = 0.10$ fixed, the variations of $\xi_v(t)$ and $N_c^\text{max}(t)$ with the active force $f_0$ exhibit a trend similar to that of the bond number $N_b(t)$ as shown in Fig.~\ref{fig_morphology}(E). Thus, the percolating network structure at $T = 0.10$ for large active force is more mechanically stable than the network structure at $T = 0.22$. The dependence on the persistence time reveals a critical active force at which $\xi_v(t)$ and $N_c^\text{max}(t)$ indicate maximum mechanical stability of the isotropic network, while activity strengths below and above this critical value lead to less stable network structures. From Figs.~\ref{fig_morphology}(G) and \ref{fig_morphology}(H) the critical active force $f_0^c$ yields $7.00$, $9.00$, and $>15.00$ for $\tau_p = 10^6$, $10^2$, and $10^0$ respectively, consistent with the values obtained in Fig.~\ref{fig_morphology}(E). Similarly, when all particles are active, the critical active force is found to be $f_0^c = 9.00$, consistent with the value obtained in Fig.~\ref{fig_morphology}(E).

\begin{figure*}
    \centering
    \includegraphics[page=1, trim=0cm 4.4cm 0cm 0cm, clip, width=1.00\textwidth]{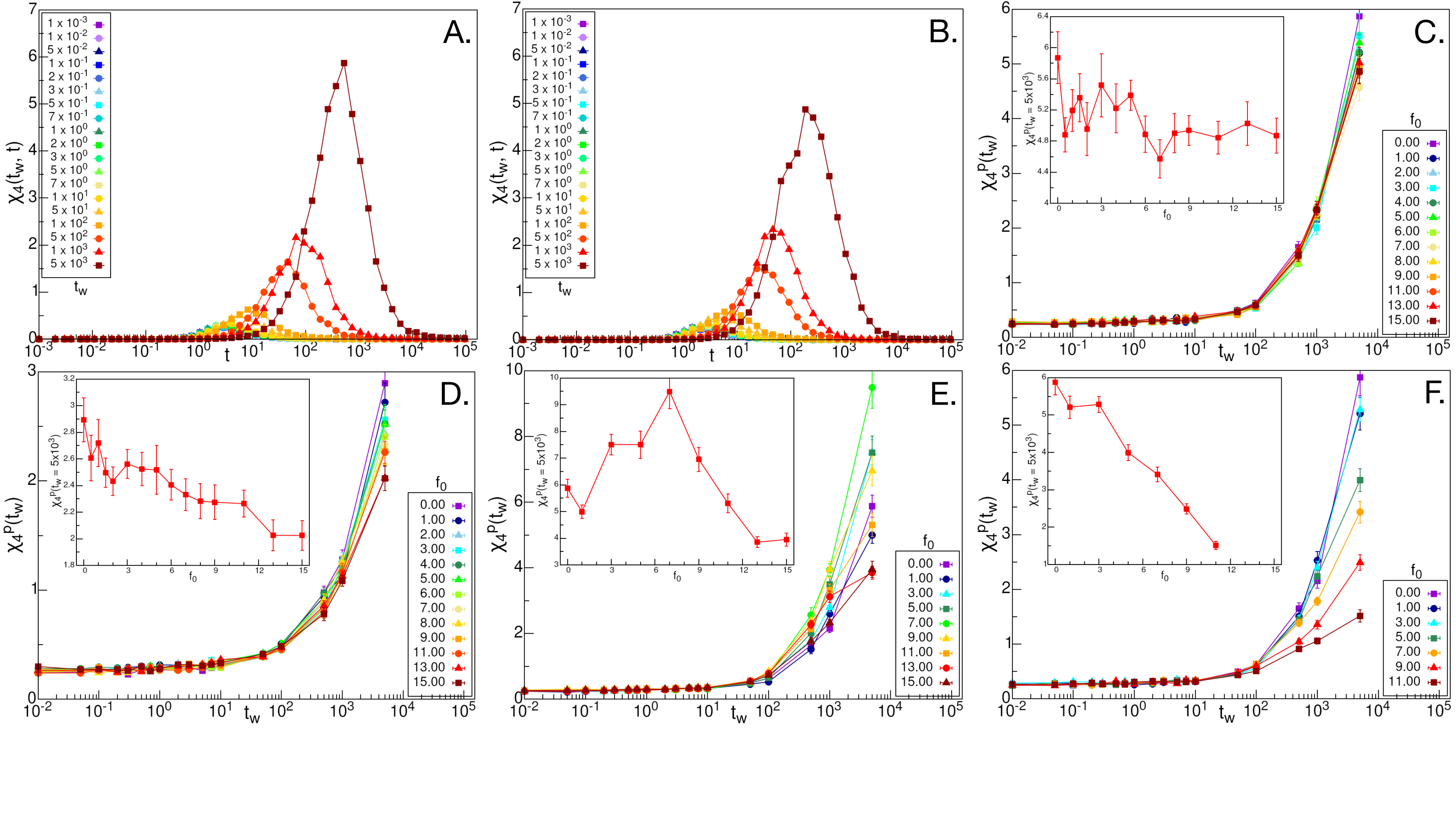}
    \caption{\textbf{Dynamical heterogeneity in aging:} Temporal evolution of $\chi_4(t_w, t)$ is shown for different waiting times for (A) $f_0 = 0.00$ and (B) $f_0 = 15.00$, with $T = 0.10$, $\tau_p = 10^0$, and $\rho_a = 0.10$ are held fixed. For a given value of $t_w$, $\chi_4(t_w, t)$ decreases with increasing $f_0$ at fixed $T = 0.10$. Waiting time dependence of the peak value of $\chi_4(t_w, t)$ is shown for the temperatures (C) $T = 0.10$ and (D) $T = 0.22$ with $\tau_p = 10^0$ and $\rho_a = 0.10$ fixed. The insets show the evolution of $\chi_4^p(t_w)$ at $t_w = 5 \times 10^3$ with increasing active force strength, indicating that $\chi_4^p(t_w)$ decreases with increasing $f_0$ at both temperatures shown in (C) and (D). (E) shows the variation of $\chi_4^p(t_w)$ with aging for different $f_0$ values is shown at a large persistence time $\tau_p = 10^6$, in comparison with (C). The inset shows the corresponding variation of $\chi_4^p(t_w)$ with $f_0$ at the $t_w = 5 \times 10^3$. For the large persistence time, $\chi_4^p(t_w)$ increases with $f_0$ up to $f_0^c \simeq 7.00$ and beyond which it begins to decrease. (F) The age-dependent growth of $\chi_4^p(t_w)$ for different $f_0$ is shown with all particles active, in comparison with panel (C). The inset shows a pronounced rapid decrease in $\chi_4^p(t_w)$ at $t_w = 5 \times 10^3$, compared with the inset of panel (C).}
    \label{fig_chi4}
\end{figure*}

\vskip+0.04in
\noindent{\bf \large Activity-dependent relaxation dynamics and dynamical heterogeneity:}
We next investigate how these structural networks influence the relaxation dynamics and dynamical heterogeneity of an active gel, characterized by the self-intermediate scattering function $F_s(K, t_w, t)$ and the block-analysis of four-point dynamical susceptibility $\chi_4(t_w, t)$ (see definition in the Model and Methods section), respectively. Figures~\ref{fig_relaxation}(A) and \ref{fig_relaxation}(B) show the time evolution of $F_s(K, t_w, t)$ for the passive ($f_0 = 0.00$) and active ($f_0 = 15.00$, $\tau_p = 10^0$, and $\rho_a = 0.10$) gels, respectively, at different waiting times, with the temperature fixed at $T = 0.10$. We evaluate $F_s(K, t_w, t)$ at a wave vector of $K = 2.50$, corresponding to a length scale smaller than the characteristic length scale associated with the $1^\text{st}$ peak of $S(K, t_w)$ for the passive gel. For a given but large $t_w$, Fig.~\ref{fig_relaxation}(A) exhibits a relatively slow and highly arrested decay of $F_s(K,t_w,t)$, reflecting the constrained particle motion imposed by the attractive and interconnected gel network, consistent with observations in fractal colloidal gels~\cite{weitz_prl_1998}. Alternatively, in the active system, enhanced particle rearrangements produce larger voids $\xi_v(t)$ and cluster size $N_c^\text{max}(t)$ in a percolating network structure, resulting in substantially faster decay. We also observe this activity-induced faster relaxation dynamics in active glasses~\cite{Paul_PNAS_2023}. To compare the wave-vector dependence of $F_s(K,t_w,t)$, we also show the relaxation behavior at a larger wave vector $K = 10.00$, in the insets of Figs.~\ref{fig_relaxation}(A) and \ref{fig_relaxation}(B). As the wave vector increases, the relaxation dynamics decay faster without an incipient plateau in both passive and active gels, compared with the small-wave-vector case. At large wave vectors, relaxation dynamics reflect fast cooperative motion of branches within the gel network, whereas at smaller wave vectors, relaxation arises from large-scale rearrangement of the overall network structure, as reported in Ref.~\cite{Gado_PRL_2007}. We also observe similar wave-vector-dependent relaxation dynamics in our active gels.

\vskip+0.04in
Next, in Figures~\ref{fig_relaxation}(C)\textendash(F), we systematically examine the age-dependent growth of the relaxation timescale $\tau(t_w)$ (see definition in the Model and Methods section) on the temperature $T$, persistence time $\tau_p$, and active particle concentration $\rho_a$ with varying active force $f_0$. Figures~\ref{fig_relaxation}(C) and \ref{fig_relaxation}(D) shows the dependence of $\tau(t_w)$ on the waiting time $t_w$ for different active force strengths at $T = 0.10$ and $0.22$, respectively, with $\tau_p = 10^0$ and $\rho_a = 0.10$ held fixed. It is evident that the relaxation time $\tau(t_w)$ remains nearly constant until the waiting time becomes comparable to the equilibrium relaxation time at the initial temperature, $t_w \leq \tau^e(T_i)$~\cite{Barrat_PRL_1997, Barrat_Phy-A_1999}. After this waiting time, the system overcomes the potential barrier of the initial temperature $T_i$ state, leading to percolating network formation and the onset of aging. Once the network starts to form, $\tau(t_w)$ grows with $t_w$, reflecting the progressive dynamical arrest of the gel and closely resembling the aging behaviour observed in colloidal Laponite suspensions in Ref.~\cite{Debasish_Soft-Matter_2014}. With increasing active force, $\tau(t_w)$ decreases at a given $t_w$ for both $T = 0.10$ and $0.22$, although the decrease is considerably smaller at the higher temperature $T = 0.22$, as shown in Fig.~\ref{fig_relaxation}(D). This difference can be directly attributed to the gelation time $t_\text{gel}$, which decreases much more rapidly with increasing activity at $T = 0.10$ than at $T = 0.22$. Thus, relaxation dynamics of an active gel are closely related to the kinetics of the gel morphology.

\vskip+0.04in
We next examine the effects of large persistent active motion $\tau_p = 10^6$ with $T = 0.10$ and $\rho_a = 0.10$, as shown in Fig.~\ref{fig_relaxation}(E). Compared with the smaller persistence motion shown in Fig.~\ref{fig_relaxation}(C), the system shows more rapid relaxation dynamics with increasing $f_0$ at fixed $ t_w$ for larger $\tau_p$. This can be attributed to the enhanced decrease in the gelation time $t_\text{gel}$ with increasing $f_0$ up to the critical active force $f_0^c = 7.00$ (see Fig.~\ref{fig_morphology}(F)), beyond which $\tau(t_w)$ no longer changes significantly at a fixed $t_w$. However, with increasing waiting time, $\tau(t_w)$ exhibits a strong enhancement up to the critical active force $f_0^c = 7.00$. Beyond this $f_0^c$, the growth of $\tau(t_w)$ with aging gradually weakens and eventually approaches a plateau, indicating a crossover from a pronounced aging regime to a stationary regime. Also, as discussed earlier, beyond $f_0^c = 7.00$, the percolating network becomes mechanically unstable. This behaviour highlights the strong connection between gel structure, mechanical stability, and relaxation dynamics. Figure~\ref{fig_relaxation}(F) shows the relaxation dynamics for a fully active system ($\rho_a = 1.00$) with $\tau_p = 10^0$ and $T = 0.10$. As in the large persistence-time regime, for a fixed $t_w$, $\tau(t_w)$ decreases rapidly with increasing $f_0$, in line with the decrease in the gelation time $t_\text{gel}$ with increasing $f_0$, as shown in Fig.~\ref{fig_morphology}(F). At the large $t_w$, increasing $f_0$ up to $f_0^c = 9.00$, systematically reduces $\tau(t_w)$, and beyond this $f_0^c$, $\tau(t_w)$ starts to approaching an steady state. Alternatively, beyond this critical active force $f_0^c$, active gels start to develop smaller void lengths and become structurally unstable, leading to enhanced fluctuations and destabilisation of the percolating network, hence faster relaxation dynamics. These results demonstrate the strong coupling between active-gel morphology and relaxation dynamics.

\vskip+0.04in
Dynamical heterogeneity (DH) is the primary cause of spatial correlations in inhomogeneous local dynamics, quantified by the peak height of the four-point susceptibility function (see definition in the Model and Methods section). Figures~\ref{fig_chi4}(A) and \ref{fig_chi4}(B) shows temporal variation of $\chi_4(t_w, t)$ for the passive gel ($f_0 = 0.00$ and $T = 0.10$) and active gel ($f_0 = 15.00$, $T = 0.10$, $\tau_p = 10^0$ and $\rho_a = 0.10$), respectively, at different waiting times. It shows that, with increasing waiting time, the peak of the four-point susceptibility $\chi_4^p$, appears at longer timescales and its magnitude increases with $t_w$. Consequently, this growth of $\chi_4^p$ during aging indicates an enhancement of DH in colloidal gels. Figure~\ref{fig_chi4}(B) shows that, upon introducing activity, the peak height of $\chi_4(t_w, t)$ is dramatically reduced at a fixed $t_w$, compared with that of the passive gel in Fig.~\ref{fig_chi4}(A). A similar behavior of $\chi_4$ has also been observed in active glasses~\cite{Paul_PNAS_2023}. Next, we show how $\chi_4(t_w, t)$ varies during aging with temperature $T$, persistence time $\tau_p$, and active particle concentration $\rho_a$, while varying the active force $f_0$. Figures~\ref{fig_chi4}(C) and \ref{fig_chi4}(D) shows the age dependent growth of $\chi_4^p$ in variation of $f_0$ for the temperatures $T = 0.10$ and $0.22$, respectively. As with the relaxation dynamics, $\chi_4^p$ starts to grow only when the waiting time exceeds the equilibrium relaxation time at the initial temperature, $t_w \ge \tau^e(T_i)$, whereas no significant DH is observed at shorter waiting times. For a given large waiting time, i.e., $t_w = 5 \times 10^3$ with zero activity, the $\chi_4^p$ attains a lower value at the higher temperature $T = 0.22$, compared to $T = 0.10$; this is a characteristic of four-point susceptibility in colloidal gels, as also observed in aging glasses~\cite{santu_arxiv_2025}. Insets of Fig.~\ref{fig_chi4}(C) and \ref{fig_chi4}(D) shows the variation of $\chi_4^p$ with active force at $t_w = 5 \times 10^3$ for the temperature $T = 0.10$ and $0.22$, respectively. As $f_0$ increases, the $\chi_4^p$ rapidly decreases and eventually saturates for $T = 0.10$, whereas at $T = 0.22$, it gradually decreases without reaching saturation over the investigated range of $f_0$. Consistently, $t_\text{gel}$ decreases rapidly with $f_0$ at $T = 0.10$, but more gradually at $T = 0.22$ (see Fig.~\ref{fig_morphology}(F)). This suggests that the extent of DH is closely related to the kinetics of gel formation and, consequently, to the evolving structure of the percolating gel network.

\vskip+0.04in
Next, the effects of persistence timescale and active particle concentration on the $\chi_4^p$ are studied by setting $\tau_p = 10^6$ and $\rho_a = 1.00$, at $T = 0.10$, in Figs.~\ref{fig_chi4}(E) and \ref{fig_chi4}(F), respectively. Similar to the earlier observation, in the large $\tau_p$ and $\rho_a$ limit, no pronounced DH is observed up to the waiting time $t_w \simeq \tau^e(T_i)$. After this $t_w$, DH grows markedly as the percolating network begins to form, suggesting that the emergence of the network structure is accompanied by an enhancement of DH. Insets of Figs.~\ref{fig_chi4}(E) and \ref{fig_chi4}(F) shows the variation of $\chi_4^p$ at the $t_w = 5 \times 10^3$, with the active force $f_0$. In the large persistence time limit, $\chi_4^p$ increases with $f_0$ up to $f_0^c \simeq 7.00$, beyond which it decreases rapidly with increasing $f_0$. Such a non-monotonic dependence of $\chi_4^p$ on $f_0$ in large $\tau_p$ limit has also been observed in active glasses~\cite{Paul_PNAS_2023, Subhodeep_Arxiv_2026}. However, for the fully active system, the $\chi_4^p$ rapidly decreases with $f_0$. The enormous decrease in $\chi_4^p$ follows a similar trend to the decrease in the gelation time $t_\text{gel}$ in the regime of large $\rho_a$.

\begin{figure*}
    \centering
    \includegraphics[page=1, trim=0cm 12.3cm 0cm 0cm, clip, width=1.00\textwidth]{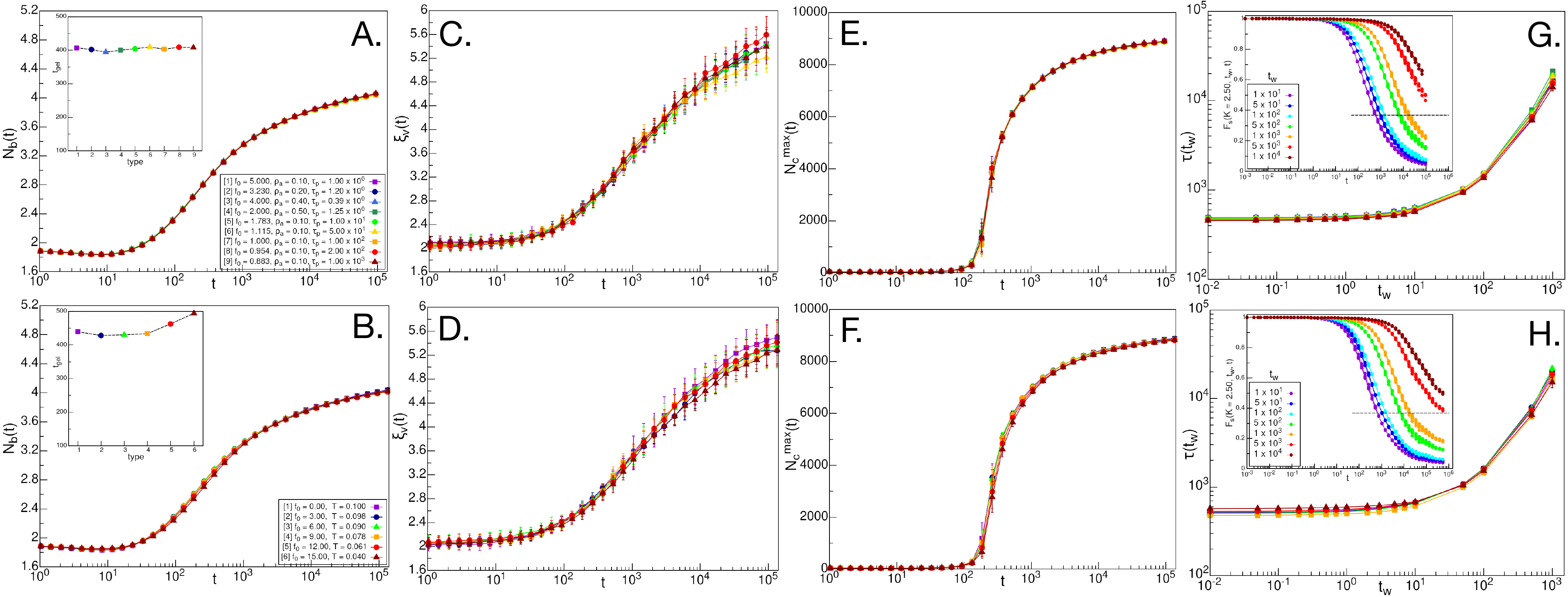}
    \caption{\textbf{Effective temperature perspective:} Equivalence of different morphological analyses and relaxation timescale at the same $T_\text{eff}$. The top panel shows results with the same $T_\text{eff}$ value obtained for different choices of $f_0$, $\tau_p$, and $\rho_a$ at $T = 0.10$, using Equ.~\ref{equ_teff_1}, while the bottom panel shows the corresponding results for different choices of $T$ and $f_0$, with $\tau_p=10^0$ and $\rho_a=0.10$ fixed, using Equ.~\ref{equ_teff_2}. (A and B) Temporal evolution of the average number of bonds per particle $N_b(t)$ is shown for different types of activity parameters. Insets show the equivalent gelation time $t_\text{gel}$ for different types of activity parameters at the same $T_\text{eff}$. (C and D) Temporal evolution of the largest void length $\xi_v(t)$ is shown for different sets of activity parameters chosen to yield the same $T_\text{eff}$. (E and F) Temporal growth of the number of isostatic particles in the largest cluster $N_c^{\max}(t)$ is shown for different activity parameters at the same $T_\text{eff}$. (G and H) The waiting time dependence of $\tau(t_w)$ is shown for different sets of activity parameters, such that $T_\text{eff}$ is held fixed. The insets show the waiting-time dependence of the decay of $F_s(K,t_w,t)$ for $t_w \in (10^1-10^4)$ with different activity parameters at the same $T_\text{eff}$.}
    \label{fig_Teff}
\end{figure*}

\vskip+0.04in
\noindent{\bf \large Effective temperature for active gels:}
Active systems are inherently non-equilibrium because they continuously consume energy to generate self-propulsion. Many propositions suggest that these non-equilibrium active particle systems can be mapped onto their equilibrium passive counterparts within an effective-temperature framework~\cite{Marjolein_Soft-Matter_2016, Cugliandolo_PRE_2008, Ludovic_EuroPhys-Letter_2015}. In the limit of a small persistence time, the effective temperature of an active system can be expressed as $T_\text{eff} = T(1 + v_0^2/2D_R D_T)$, where $v_0=f_0D_T/K_B T$ is the swim velocity~\cite{Marjolein_Soft-Matter_2016}. The effective temperature $T_\text{eff}$ is also commonly called the swim temperature~\cite{Claudio_Soft-Matter_2015}. In experiments~\cite{Bocquet_PRL_2010}, this $T_\text{eff}$ is estimated using the fluctuation-dissipation relation $K_B T_\text{eff} = D_\text{eff}/\mu$, by measuring the particle diffusion $D_\text{eff}$ and mobility $\mu$ under gravity. Following Refs.~\cite{Saroj_PNAS_2018, Paul_PNAS_2023}, we define the activity induced excess temperature as $K f_0^2 \tau_p \rho_a/(1 + G\tau_p)$, which, together with the bath temperature $T$, gives the effective temperature:
\begin{equation}
    T_\text{eff} = T + K \times \frac{f_0^2 \tau_p \rho_a}{1 + G\tau_p}.
    \label{equ_teff_1}
\end{equation}
Here, $K$ and $G$ are being two constants. To quantify the excess temperature arising from self-propelled particles in our active gel system, we first vary $f_0$, $\rho_a$, and $\tau_p$ while keeping the bath temperature $T = 0.10$ unchanged. Next, we fix the persistence time $\tau_p = 10^0$ and active particle concentration $\rho_a = 0.10$, and vary $T$ and $f_0$ that yields the effective temperature relation,
\begin{equation}
    T_\text{eff} = T + \tilde{K} f_0^2.
    \label{equ_teff_2}
\end{equation}
Here, $\tilde{K}$ is another constant that depends on $K$ and $G$. In Fig.~\ref{fig_Teff} we show the bond number $N_b(t)$, gelation time $t_\text{gel}$, void size $\xi_v(t)$, and cluster size $N_c^\text{max}(t)$ as structural quantities and the relaxation time scale $\tau(t_w)$ as a dynamical quantity, which suggests that the effective temperature framework applies to an active gel quite well.

\vskip+0.04in
Top panel of Fig.~\ref{fig_Teff} shows the variation of structural and dynamical quantities with variation of $f_0 \in (0.883 - 5.000)$, $\rho_a \in (0.10 - 0.50)$, and $\tau_p \in (3.90 \times 10^{-1} - 10^3)$, chosen such that the quantity $\frac{f_0^2 \tau_p \rho_a}{1 + G\tau_p} \sim 2.43$ remains approximately constant, with $G = 0.03125$. $G$ is also found to be a constant in an active glass, as reported for systems simulated using active Ornstein--Uhlenbeck particles~\cite{Puneet_PNAS_2026} and run-and-tumble particles~\cite{Subhodeep_PRE_2025}. Next, the bottom panel of Fig.~\ref{fig_Teff} shows a similar variation of the structural and dynamical quantities for different choices of $f_0$ and $T$, chosen such that $T_\text{eff} = 0.10$, thereby mimicking the behavior of an active gel by its passive counterpart. We obtained $\tilde{K} = 2.70 \times 10^{-4}$, thus by setting $\tau_p = 10^0$ and $\rho_a = 0.10$, in the relation $\tilde{K} = K \times \frac{\tau_p \rho_a}{1 + G\tau_p}$, yields the parameter $K = 2.78 \times 10^{-3}$.

\begin{figure*}
    \centering
    \includegraphics[page=1, trim=0cm 9.5cm 0cm 0cm, clip, width=1.00\textwidth]{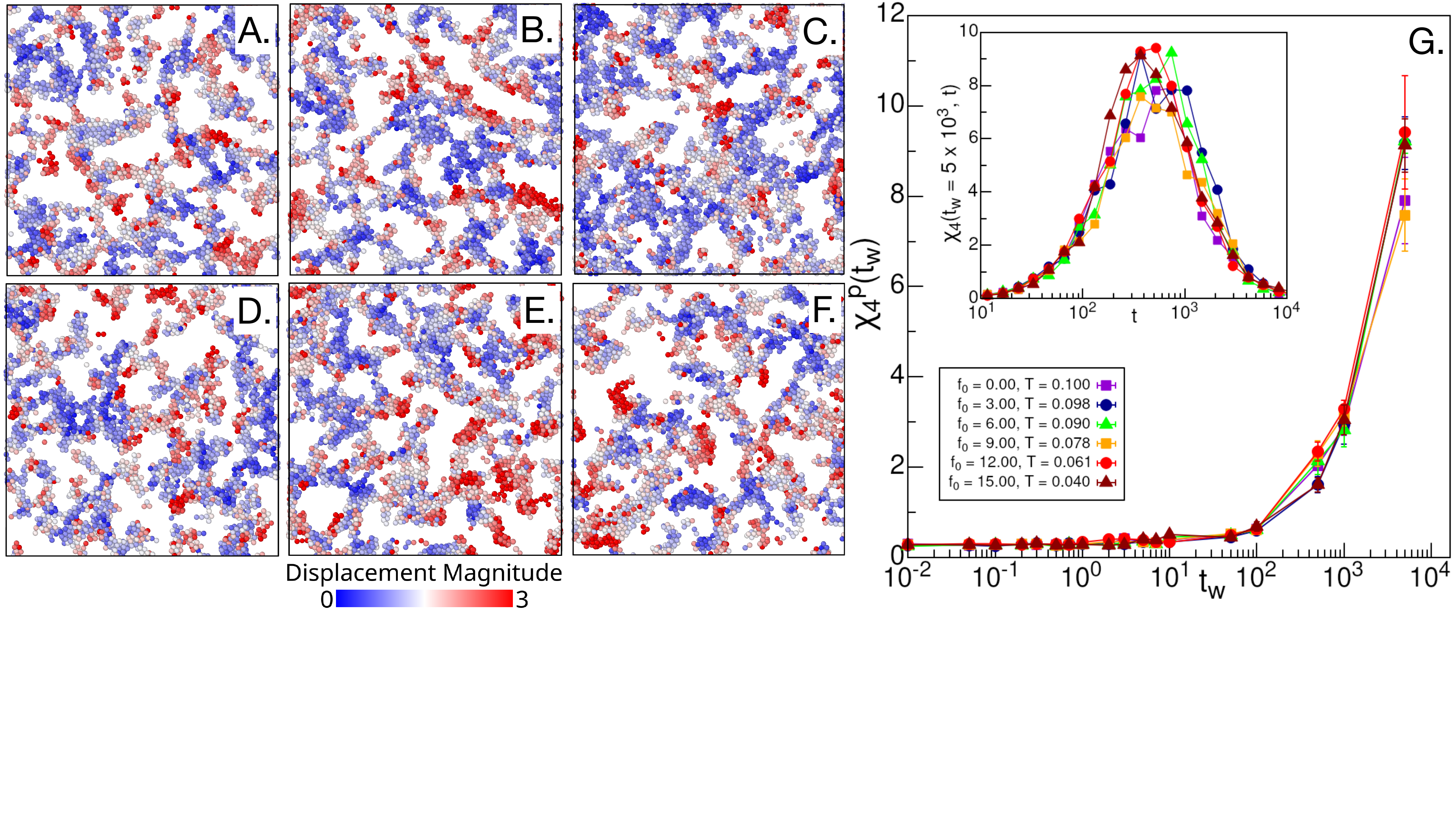}
    \caption{\textbf{Structural and dynamical heterogeneity:} Particle displacement in the percolating network at $t \sim 9.7 \times 10^4$, measured with respect to the corresponding gelation time $t_\text{gel} \sim 470$, for different combinations of temperature and active force: (A) $T = 0.100$, $f_0 = 0.00$; (B) $T = 0.098$, $f_0 = 3.00$; (C) $T = 0.090$, $f_0 = 6.00$; (D) $T = 0.078$, $f_0 = 9.00$; (E) $T = 0.061$, $f_0 = 12.00$; and (F) $T = 0.040$, $f_0 = 15.00$, chosen such that all parameter sets have the same $T_\text{eff}$. The color legend indicates the corresponding particle displacement magnitude, with blue representing particles that overlap with their initial positions and red representing particles displaced by up to three particle diameters. Across panels (A)\textendash(F), the gel networks exhibit similar void and cluster sizes, along with nearly identical spatial distributions of mobile and immobile particles. (G) The age-dependent growth of $\chi_4^p(t_w)$ is shown for the same sets of $T$ and $f_0$ with $N = 50\,000$, indicating similar dynamical heterogeneity across all cases. The inset shows the temporal evolution of $\chi_4(t_w,t)$ at $t_w = 5 \times 10^3$, illustrating similar DH over time at this fixed $t_w$.}
    \label{fig_sh_vs_dh}
\end{figure*}

\vskip+0.04in
Figure~\ref{fig_Teff}(A) shows qualitatively similar temporal evolution of $N_b(t)$ for different combinations of $f_0$, $\rho_a$, and $\tau_p$ with $T = 0.10$ fixed, which indicates that despite the large variations in activity parameters, the resulting percolating networks possess comparable morphologies. The inset shows the corresponding gelation time $t_\text{gel}$ for different activity parameters, indicating a nearly identical timescale for percolating network formation. Next, we vary $f_0$ from $0.00$ (passive) to $15.00$ (strongly active), while adjusting the corresponding temperature $T$ from $0.100$ to $0.040$, with $\rho_a = 0.10$ and $\tau_p = 10^0$ held fixed, such that $T_\text{eff} = 0.10$ is maintained. The main panel and inset of Fig.~\ref{fig_Teff}(B) show the temporal evolution of $N_b(t)$ and $t_\text{gel}$, respectively. This illustrates that within the $T_\text{eff}$ framework, increasing $f_0$ yields an active gel with structural connectivity and gelation kinetics comparable to those of the passive gel. Figure~\ref{fig_Teff}(C) shows a qualitatively similar temporal evolution of the void length $\xi_v(t)$ for different choices of activity parameters, suggesting that the characteristic pore structure of the percolating gel networks remains essentially unchanged despite variations in the activity parameters within the $T_\text{eff}$ framework. Similarly, in Fig.~\ref{fig_Teff}(D), we show temporal evolution of void size $\xi_v(t)$ in variation with $f_0$ such that $T_\text{eff} = 0.10$. This shows that $\xi_v(t)$ exhibits qualitatively similar temporal evolution in the active system and its passive counterpart, indicating that the characteristic pore structure remains largely unchanged despite variations in the strength of the active force $f_0$ within the $T_\text{eff}$ framework. Next, we quantify the temporal evolution of the number of isostatic particles in the largest percolating network $N_c^\text{max}(t)$ as a measure of mechanical stability from an effective-temperature perspective. Figures~\ref{fig_Teff}(E) and \ref{fig_Teff}(F) show qualitatively similar temporal growth of $N_c^\text{max}(t)$ for different choices of activity parameters and with increasing $f_0$, respectively, given that the $T_\text{eff}$ relationship holds. This implies that, within the $T_\text{eff}$ framework, active gels with different activity parameters exhibit comparable mechanical stability and, as $f_0$ increases, mechanical stability is preserved, essentially equivalent to that of the corresponding passive gel. All results in Figs.~\ref{fig_Teff}(A)–(F) suggest that, within the $T_\text{eff}$ framework, the gel network's structural integrity and mechanical stability remain essentially unchanged.

\vskip+0.04in
We now demonstrate the age-dependent growth of relaxation dynamics $\tau(t_w)$ within the effective temperature framework. Figure~\ref{fig_Teff}(G) shows similar exponential growth of $\tau(t_w)$ with waiting time for all active systems with varying activity parameters, further validating our effective-temperature analogy, as previously demonstrated for active glasses~\cite{Paul_PNAS_2023}. Inset shows the corresponding time evolution of $F_s(K, t_w, t)$ for $t_w \in (10^1 - 10^4)$. It demonstrates that not only is $\tau(t_w)$ equal, but also the temporal evolution of $F_S (K, t_w, t)$ is similar for all choices of activity parameters that yield the same $T_\text{eff}$. Figure~\ref{fig_Teff}(H) shows that active systems with different activity strengths exhibit a similar age-dependent growth of $\tau(t_w)$ and similar temporal evolution of $ F_S (K,t_w,t)$ as their passive counterparts, further supporting the validity of the effective temperature framework.

\vskip+0.04in
\noindent{\bf \large Structural and dynamical heterogeneity in active gels:}
To this point, we have seen that activity can alter the structural and dynamical properties of colloidal gels, which have also been reported from different perspectives in Refs.~\cite{chandana_prl_2024, brady_pnas_2024, olivier_prl_2023}. In an active system, local energy injection through active-particle self-propulsion can restructure the colloidal gel network into a highly porous medium~\cite{chandana_prl_2024}. Moreover, Ref.~\cite{olivier_prl_2023} shows that photo-switchable active doping of Janus particles in a polystyrene gel network enhances particle mobility while preserving the network connectivity, resulting in a softer gel with arrested aging dynamics. Also, Ref~\cite{brady_pnas_2024} shows that activity can alter the mechanical response of the network of colloidal gels. Following the results discussed earlier, an active gel can often be characterized by an effective temperature, and within the $T_\text{eff}$ description, the structure and relaxation timescale of an active gel are qualitatively similar to those of its passive counterpart. However, within the $T_\text{eff}$ description, an active glass can exhibit relaxation timescales comparable to their equilibrium counterparts, but activity can have nontrivial effects on their dynamical heterogeneity~\cite{Paul_PNAS_2023}. We are now seeking to investigate in active gels whether the dynamical heterogeneity exhibits a nontrivial growth or whether its behavior is fundamentally different from that observed in active glasses.

\vskip+0.04in
To address this, we now compare the structures and dynamical heterogeneity of different active gels at the same effective temperature $T_\text{eff}$. Figures~\ref{fig_sh_vs_dh}(A\textendash F) shows the representative cross-sectional snapshots at the time $t \sim 9.7 \times 10^4$, with different active forces at $T_\text{eff} = 0.10$. All representative snapshots show similar void and cluster sizes, consistent with the similar values of $\xi_v(t)$ and $N_c^\text{max}(t)$ quantified in the previous section. This indicates that all configurations exhibit similar structural heterogeneity, though the active force strengths differ. The blue and red colours indicate slow and fast moving particles, with displacements ranging from zero to three particle diameters. We measure these particle displacements from the gelation time $t_\text{gel} \sim 470$. The populations of slow (blue) and fast (red) particles are similar across all configurations, indicating comparable spatial heterogeneity of particle mobility. Figures~\ref{fig_sh_vs_dh}(A\textendash F) show that these representative active-gel configurations have similar structural and spatial heterogeneity and the same relaxation timescale. This similarity in spatial heterogeneity of particle mobility is further confirmed by the Van Hove function, as shown in Appendix D.

\vskip+0.04in
Next, we compute $\chi_4(t_w, t)$ to quantify the temporal evolution of spatial heterogeneity with increasing $f_0$, at a similar $\tau(t_w)$ for $N = 50\,000$. Figure~\ref{fig_sh_vs_dh}(G) shows that $\chi_4^p(t_w)$ exhibits similar growth with waiting time for systems that have comparable age-dependent growth of $\tau(t_w)$. The insets of Fig.~\ref{fig_sh_vs_dh}(G) show the temporal evolution of $\chi_4(t_w, t)$ at $t_w = 5 \times 10^3$, indicating that the dynamical heterogeneity evolves qualitatively similarly over time for those active gels with different activity strengths $f_0$, despite having similar $\tau(t_w)$. These results demonstrate that, despite having different activity strengths $f_0$ at the same $T_\text{eff}$, the systems exhibit similar structural and dynamical heterogeneity. This indicates that, in colloidal gels, competition between activity and temperature produces similar structures at the same $T_\text{eff}$. Thus, dynamical heterogeneity in active gels with highly heterogeneous structure differs markedly from that observed in active glasses with a homogeneous structure.

\section{Conclusion}
In conclusion, we have conducted extensive Brownian dynamics simulations of active gels to investigate the influence of activity on the structural and dynamical evolution of the percolating gel network. We show that introducing active agents transforms the ramified network into a well-annealed, mechanically stronger structure, with faster gelation kinetics and larger void and cluster sizes. Essentially, active doping promotes the formation of well-annealed percolating network structures, particularly at lower temperatures ($T = 0.10$), indicating that active particles act as local structural reorganization agents, enabling the system to explore lower-energy configurations, consistent with the findings in Ref.~\cite{omar_acs-nano_2018}. We also demonstrate that an optimal swim force exists for engineering well-annealed gel network structures, and beyond this critical swim force, the gel network becomes destabilized. This critical swim force strongly depends on the persistence timescale $\tau_p$ and the active particle concentration $\rho_a$.

\vskip+0.04in
Conversely, activity alters the dynamics of colloidal gels during aging, as quantified by the evolution of the structural relaxation timescale and dynamical heterogeneity for different activity parameters. Our results show that, particularly at low temperature, the percolating network structures relax faster and become less heterogeneous with increasing swim force than at higher temperature ($T = 0.22$). This suggests that activity enhances particle rearrangements, leading to the accumulation of mobile particles into larger clusters, which reduces the spatial correlations between the bimodal populations of mobile and immobile particles, thereby decreasing the dynamical heterogeneity. Further, we show that increasing $\tau_p$ and $\rho_a$ reduces the relaxation timescale up to a critical swim force $f_0^c$. Beyond this critical swim force, the percolating network structure becomes destabilized, and the corresponding relaxation timescale leaves the aging regime and reaches a stationary state. Collectively, these findings indicate that the structure of colloidal gels is strongly influenced by the competition between thermal fluctuations characterized by $T$ and activity parameters that are characterized by $f_0$, $\tau_p$, and $\rho_a$, which in turn govern the relaxation timescale and dynamical heterogeneity of the colloidal gels.

\vskip+0.04in
Most interestingly, the effective temperature ($T_\text{eff}$) framework turns out to be a good descriptor for these active colloidal gels, mapping several structural quantities, such as the bond number, void size, and cluster size of active systems onto those of their passive counterparts. On the dynamical aspects, our results also validate the equivalence of relaxation timescales between active and passive colloidal gels within the same $T_\text{eff}$ framework, which has been widely established for model active glasses~\cite{Paul_PNAS_2023} and glassy dynamics in dense cellular tissues~\cite{Sussman_PRE_2025}. However, in glasses, even when the relaxation timescale is similar, the DH still grows non-trivially with swim force. In contrast, in active colloidal gels, the well-debated growth of DH with swim force is similar for systems with comparable relaxation timescales. This result highlights a fundamental difference between gels and glasses\textemdash in colloidal gels, structural heterogeneity plays an important role in determining DH, whereas glasses have homogeneous structures, which do not control the DH of glasses. Recently, Ref.~\cite{Subhodeep_Arxiv_2026} reports a non-trivial structure-dependent DH in two-dimensional active glasses exhibiting medium-range crystalline order (MRCO). This suggests that although the effective temperature framework describes relaxation times in active materials quite well in a universal manner, dynamical heterogeneity might depend in an extremely non-trivial way on activity, which remains unexplored systematically.

\vskip+0.04in
The validity of an effective temperature description in active gels represents a significant advancement, as it may enable the tuning of gel properties for future industrial applications. Our findings highlight the essential role of structural heterogeneity in governing dynamical heterogeneity in colloidal gels and demonstrate that the strong coupling between these heterogeneities distinguishes colloidal gels from conventional glass-forming liquids. We anticipate that this work will encourage further experimental studies to systematically examine the interplay between structural and dynamical heterogeneity in active colloidal gels across a broad range of activity parameters. In future research, it will be important to elucidate the effects of activity during the transition from gels to glasses by systematically increasing system density. Such investigations may clarify how the decoupling between dynamical heterogeneity and structure occurs as systems approach glassy state from gels. Ultimately, we believe that our findings offer valuable insights into how activity can be harnessed to engineer colloidal gels with broader and more robust percolating network strands for societal benefit.

\section{Model and Methods}
\noindent{\bf \large Model details:} 
We have studied a well-constructed model of a size-polydisperse colloid-polymer mixture~\cite{benjamin_jour-of-rheo_2014, benjamin_jour-of-rheo_2016, himangsu_pre_2025, himangsu_soft-matt_2025}, interacting via a short-range strongly attractive Morse potential
\begin{equation}
    U(r) = 
\begin{cases}
   \epsilon_0 \left[e^{-2\alpha_0(r - \sigma_{ij})} - 2e^{-\alpha_0(r - \sigma_{ij})} \right], \hspace{0.2cm} r<r_c, \\
   0, \hspace{4.9cm} r\ge r_c.
\end{cases}
\label{equ_morse}
\end{equation}
In our model system, particles have a continuous size distribution comprising seven particle species with diameters ranging from $0.76 \bar{\sigma}$ to $1.24 \bar{\sigma}$. We choose the relative concentrations of the different particle species to reproduce a Gaussian distribution with $8\%$ polydispersity, setting the average particle diameter, $\bar{\sigma}$, to unity. The depth of the potential is $\epsilon_0 = 1$, the width of the potential is $\alpha_0 = 33 \bar{\sigma}^{-1}$, the average diameter of interacting particles $i$ and $j$ is $\sigma_{ij} = 0.5(\sigma_i + \sigma_j)$, and the cutoff parameter is $r_c = 1.4 \sigma_{ij}$. The length, energy, and time units are given by $\bar{\sigma}$, $\epsilon_0$, and $\sqrt{m \bar{\sigma}^2 / \epsilon_0}$.

\vskip+0.04in
\noindent{\bf \large Activity modeling:} 
We model activity in our colloidal gel system as active Brownian particles, simulated using Langevin dynamics~\cite{omar_acs-nano_2018, mallory_jcp_2020, fergus_jcp_2023, brady_pnas_2024, chandana_prl_2024}. We implement the activity by assigning a self-propulsion force $f_0$ to a fraction $ \rho_a$ of randomly chosen particles. Each active particles experience a propulsion force $\Vec{F^a_i}(t) = f_0 \Vec{n_i}(t)$, where $\Vec{n_i}(t)$ is an unit vector defining the propulsion direction. The three-dimensional translational and rotational diffusion motion of a particle is governed by the Langevin dynamics,
\begin{equation}
    m \frac{d^2\Vec{r_i}}{dt^2} = -\frac{\partial U(r)}{\partial \Vec{r_i}} - \gamma_0 \frac{d \Vec{r_i}}{dt} + \Vec{F_i^a}(t) + \sqrt{2\gamma_0^2D_T} \, \Vec{\eta}(t),
    \label{equ_translation}
\end{equation}
\begin{equation}
    \frac{d \vec{n_i}}{d t} = \sqrt{2D_R} \, \big(\vec{\xi_i}(t) \times \vec{n_i}(t)\big).
    \label{equ_rotation}
\end{equation}
Here, $\Vec{r_i}$ is the position vector of the $i^\text{th}$ particle, and $m$ is the mass of the particle. We choose the friction coefficient $\gamma_0 = 100$, corresponding to the overdamped regime, where viscous drag dominates over inertial effects, as appropriate for a colloid-polymer mixture~\cite{zottl_annu-rev_2023}. The term $\vec{\eta}(t)$ describes zero-mean Gaussian white noise that promotes thermal fluctuations in translational motion. The translational diffusion coefficient $D_T = K_B T/\gamma_0$, where $K_B$ is the Boltzmann constant, and $T$ is the temperature. In Equ.~\ref{equ_rotation}, $\vec{\xi_i}(t)$ is an independent Gaussian noise term that drives the active particle in the direction of $\vec{n_i}(t)$, and $D_R$ is the rotational diffusion coefficient, which is characterized by the persistence time $\tau_p = 1/2D_R$~\cite{adam_epl_2014, fergus_jcp_2023}.

\vskip+0.04in
\noindent{\bf \large Simulation details and protocols:} 
The equations of motion (Equs.~\ref{equ_translation} and \ref{equ_rotation}) are integrated using the BAOAB integrator~\cite{benedict_applied-math_2013} with a time step of $dt = 10^{-3}$. Our model system consists of $N = 10\,000$ particles confined in a cubic simulation box with periodic boundary conditions in all directions. Simulations are additionally performed for a larger system of $N = 50\,000$ particles for visualization of snapshots. We fix the packing fraction at $\phi = 0.20$. Following Refs.~\cite{patrick_prl_2015, patrick_jcp_2017, patrick_soft-matt_2017, Patrick_JCP_2018}, we choose the simulation temperature to satisfy the criterion for forming a percolating gel network, $\epsilon_0/K_BT \ge 3.13$ (see Appendix A). An aging state in a colloidal-gel system has been achieved by initially making a percolating fluid at temperature $T_i = 1.00$, and subsequently quenching the system at time $t = 0$ to a final temperature $T = 0.22$ and $0.10$. Simulations have been performed by varying $f_0 \in (0.00 - 15.00)$, $\rho_a \in (0.10 - 1.00)$ and $\tau_p \in (10^0 - 10^6)$ at the temperature $T$. Additionally, simulations within the effective temperature framework are performed at fixed $T = 0.10$, by varying $f_0$, $\rho_a$, and $\tau_p$. Simulations are also carried out for $T_\text{eff} = 0.10$ for different combinations of $T$ and $f_0$ with $\rho_a = 0.10$ and $\tau_p = 10^0$ held fixed. All analyses have been performed after evolving at $T$ for a waiting time $t_w \in (10^{-3} - 10^5)$. All simulations have been performed using an in-house OpenMP Fortran code with $16$ statistically independent ensembles for averaging.

\vskip+0.04in
\noindent{\bf \large Dynamical correlation functions:}
The non-stationary variant of the two-point time correlation function in reciprocal space (self-intermediate scattering function)~\cite{Barrat_PRL_1997,Barrat_Phy-A_1999} for a system of $N$-point particles is defined as,
\begin{equation}
    F_s(K,t_w, t) = \Bigg[\frac{1}{N} \sum_{j = 1}^{N} e^{-i\Vec{K}\cdot \big(\Vec{r_j}(t_w+t) - \Vec{r_j}(t_w)\big)} \Bigg],
    \label{equ_fskt}
\end{equation}
where, the term $\big(\Vec{r_j}(t_w+t) - \Vec{r_j}(t_w)\big)$ denotes the displacement vector of the $j^\text{th}$ particle over a time $t$, measured from the waiting time $t_w$. $[...]$ denotes the average over multiple realizations, and an additional average is taken over different isotropic directions of $\vec{K}$. We choose the wave-vector magnitude $|\vec{K}|$ from the $1^\text{st}$ peak of the static structure factor $S(K, t_w)$ (see Appendix B). Unless stated otherwise, $F_s(K, t_w, t)$ is computed for the value of the wave vector $K = 2.50$. The structural relaxation time $\tau(t_w)$ characterizing the dynamics of the percolating gel, at different waiting times $t_w$, is estimated by setting $\big[F_s\big(K, t_w, t = \tau(t_w)\big)\big] = 1/e$~\cite{Barrat_PRL_1997, Barrat_Phy-A_1999}.

\vskip+0.04in
During the aging of a percolating gel, the network structure undergoes slow structural reorganization, leading to the coexistence of slow and fast moving regions, which gives rise to an increase in the dynamical heterogeneity (DH). We quantify the age-dependent growth of DH using the four-point susceptibility $\chi_4$~\cite{dasgupta_europhys-lett_1991}, which measures fluctuations in the overlap function $q_s$ (see Appendix C). Calculation of $\chi_4$ in out-of-equilibrium dynamics requires averaging over a large number of realizations. However, block-analysis~\cite{Saurish_PRL_2017} offers an efficient and experimentally feasible method for estimating $\chi_4$ by averaging over smaller sub-systems of a larger system. We divide a system of linear size $L$ into smaller sub-systems, each of linear size $L_B = L/n$, where $n \in \{3, 4, 5 ... \}$. The nonstationary variant of block-analysis of the self-overlap function and the corresponding dynamic susceptibility~\cite{santu_arxiv_2025} for a system of $N$-point particles is defined as,
\begin{align}
    \notag
    & q_s(L_B, t_w, t) = \\
    & \hspace{1cm} \frac{1}{N_B} \sum_{i=1}^{N_B} \frac{1}{n_i(t_w)} \sum_{j=1}^{n_i(t_w)} w\Big(\big|\Vec{r_j}(t_w + t) - \Vec{r_j}(t_w)\big|\Big), \\
    \notag
    & \chi_4(L_B,t_w,t) = \\
    & \hspace{1cm}  \frac{N}{N_B}  \left[\Big<q_s^2(L_B,t_w,t)\Big> - \Big< q_s(L_B,t_w,t) \Big>^2 \right],
\end{align}
where, $N_B$ is the number of blocks with size $L_B$, $n_i(t_w)$ is the number of particles in the $i$-th block at $t_w$, and $w(x) = \Theta(a - x)$ is the window function, with $\Theta$ being the Heaviside step function. The parameter ``$a$'' is chosen from the plateau height of the mean-squared displacement (see Appendix C). $\langle...\rangle$ and $[...]$ denote averaging over different blocks of size $L_B$ and different realizations, respectively. We calculate $\chi_4(L_B,t_w,t)$ for $L_B = L/3$ and denote it as $\chi_4(t_w,t)$ in the main text, since $L_B$ is fixed throughout this work.

\section{Acknowledgment}
We acknowledge the funding from intramural funds at TIFR Hyderabad from the Department of Atomic Energy (DAE) under Project Identification No. RTI 4007. We thank Alessio Zaccone for giving valuable feedback on our manuscript. SK acknowledges the Swarna Jayanti Fellowship grants DST/SJF/PSA01/2018-19 and SB/SFJ/2019-20/05 from the Department of Science and Technology (DST) and the Science and Engineering Research Board (SERB). SK acknowledges the J C Bose Grant ANRF/JBG/2026/000149/PS from the Anusandhan National Research Foundation (ANRF), Government of India. SK acknowledges research support from the MATRICES Grant MTR/2023/000079, funded by SERB. Most computations were performed on HPC clusters procured with Swarna Jayanti Fellowship grants and the Core Research Grant CRG/2019/005373.

\bibliographystyle{apsrev4-2}
\bibliography{bibliog.bib}

\section*{Appendix A: Onset of Gelation}
\begin{figure*}
    \centering
    \includegraphics[page=1, trim=0cm 4.2cm 0cm 0cm, clip, width=1.00\textwidth]{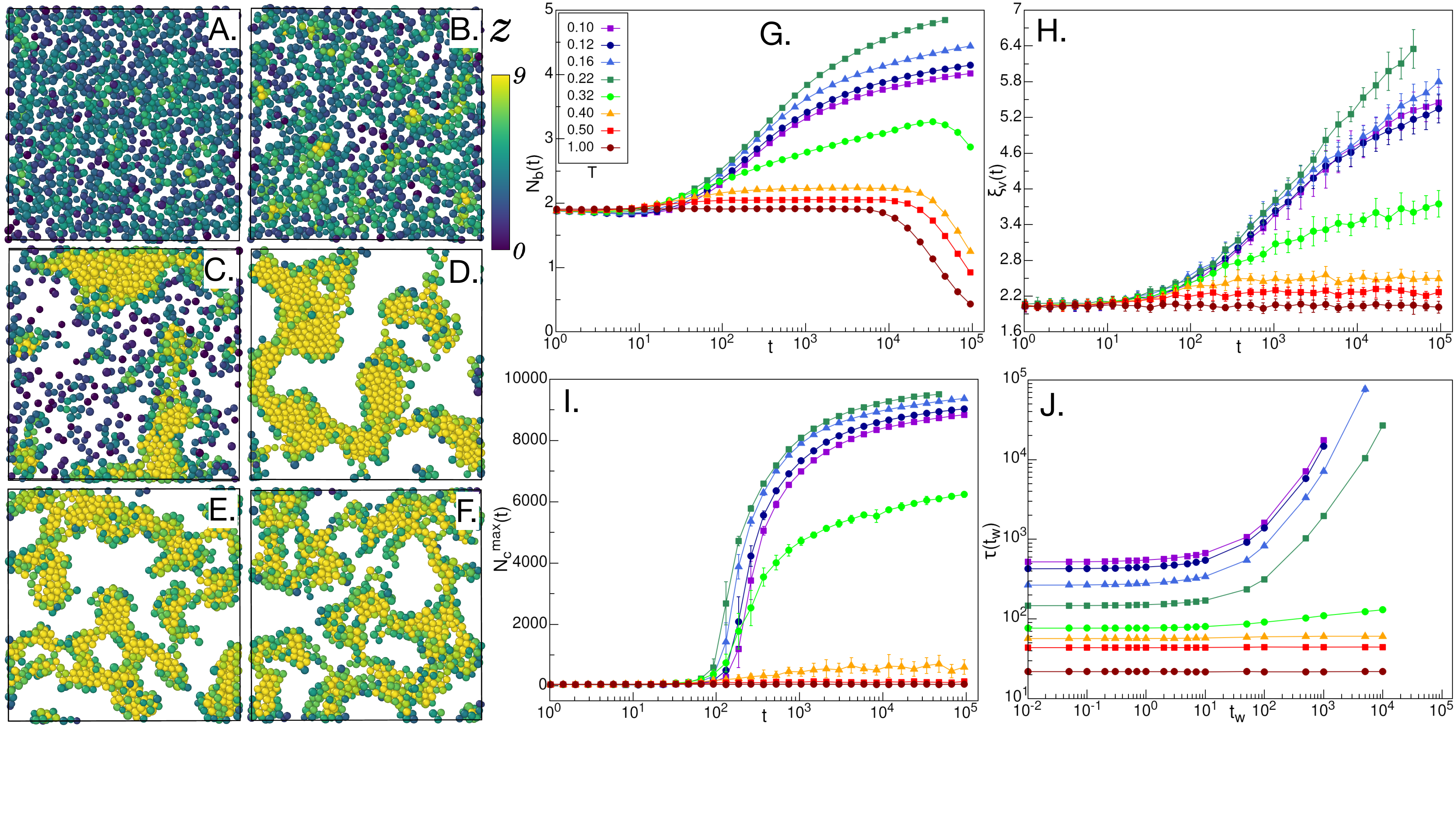}
    \caption{The representative cross-sectional snapshots of the attractive passive colloidal system at different temperatures\textemdash (A) $1.00$, (B) $0.40$, (C) $0.32$, (D) $0.22$, (E) $0.12$, and (F) $0.10$. As the temperature $T$ decreases, colloidal systems form a percolating network, leading to the formation of a gel phase. Blue and yellow particles represent coordination numbers $z = 0$ and $9$, respectively, as indicated by the color bar. Time evolution of (G) bond number $N_b(t)$, (H) void length $\xi_v(t)$, and (I) the number of isostatic particles in the largest cluster $N_c^\text{max}(t)$, for different temperatures as defined in the legend of (G). (J) shows the growth of $\tau(t_w)$ with aging time for the same temperatures.}
    \label{fig_appendix_a}
\end{figure*}

In this section, we examine the temperature-driven fluid-to-gel phase transition in our model system, focusing on the evolution of the structure and dynamics, and the emergence of arrested spinodal decomposition~\cite{Zaccarelli_Nature_2008, Patrick_Jour-of-Phys_2021, Patrick_JCP_2018}. Figures~\ref{fig_appendix_a}(A\textendash F) show cross-sectional views of the colloidal system at $t \sim 1 \times 10^5$, with the temperature decreasing from $T = 1.00$ to $0.10$. Figures~\ref{fig_appendix_a}(A) and \ref{fig_appendix_a}(B) show fully fluid-like structures with no discernible clusters of isostatic particles at $T = 1.00$ and $0.40$, respectively. The configuration at $T = 0.32$ in Fig.~\ref{fig_appendix_a}(C) shows the emergence of clusters of isostatic particles, indicating the coexistence of fluid and gel phases. Thus, in our model system, gelation begins at $T_\text{onset} = 0.32$, corresponding to an interaction strength of $\epsilon_0/K_B T = 3.125$. This result is consistent with the Noro-Frenkel criterion~\cite{Frenkel_JCP_2000}, as reported in Refs.~\cite{himangsu_pre_2025, himangsu_soft-matt_2025}, which predicts gelation or spinodal decomposition at $\epsilon_0/K_B T \simeq 3.10$. Figures~\ref{fig_appendix_a}(D\textendash F) show the gel phase at $T = 0.22$, $0.12$, and $0.10$, characterized by wider to thinner percolating network structures. We interpret this progressive structural evolution in terms of thermally activated bond-breaking processes, using the Kramers’ escape rule~\cite{Kramer_Physica_1940} as a qualitative framework, rather than as a quantitatively valid high-barrier approximation, as discussed in Ref.~\cite{Zaccone_PRE_2017}.

\vskip+0.04in
The onset of gelation is accompanied by dramatic structural changes with decreasing temperature. Having characterized the structural features of these phases at $t \sim 1 \times 10^5$, we now examine their temporal evolution by tracking the bond number $N_b(t)$ in Fig.~\ref{fig_appendix_a}(G), void length $\xi_v(t)$ in Fig.~\ref{fig_appendix_a}(H), and number of isostatic particles in the largest cluster $N_c^\text{max}(t)$ in Fig.~\ref{fig_appendix_a}(I). In Fig.~\ref{fig_appendix_a}(G), $N_b(t)$ increases progressively with time for $T \leq 0.32$, indicating the growth of an interconnected network through the development of attractive interparticle bonds. At $T = 0.32$, $N_b(t)$ reaches $3$, corresponding to an average coordination number of $z = 6$. Beyond this point, $N_b(t)$ decreases in the long-time limit, indicating the formation of an unstable network. However, at higher temperatures ($T > 0.32$), $N_b(t)$ remains below $3$, indicating that thermal fluctuations increasingly compete with attractive interactions, suppressing persistent bond formation and keeping the system predominantly in the fluid state. The corresponding temporal evolution of the void length $\xi_v(t)$ and cluster size $N_c^\text{max}(t)$, shown in Figs.~\ref{fig_appendix_a}(H) and \ref{fig_appendix_a}(I), further supports this structural picture. For $ T> T_\text {onset} $, $\xi_v(t)$ and $N_c^\text{max}(t)$ show no apparent evolution with time, whereas for $T < T_\text{onset}$ both quantities grow pronouncedly and persist at long times.

\vskip+0.04in
Figure~\ref{fig_appendix_a}(J) shows the waiting time dependence of $\tau(t_w)$ at different temperatures, revealing a clear temperature-dependent crossover in the relaxation dynamics. At high temperatures ($T > T_\text{onset}$), $\tau(t_w)$ remains nearly independent of the waiting time, indicating negligible aging effects and relatively stationary relaxation dynamics. At high temperatures, thermal fluctuations are strong enough to overcome attractive interactions that would otherwise prevent dynamical arrest. At $T = T_\text{onset}$, $\tau(t_w)$ shows weak growth at large $t_w$, indicating the onset of spinodal decomposition. As the temperature is further decreased below $T_\text{onset}$, $\tau(t_w)$ grows progressively faster with $t_w$, reflecting enhanced aging and the emergence of dynamical arrest associated with spinodal decomposition.

\begin{figure}
    \centering
    \includegraphics[width=0.46\textwidth]{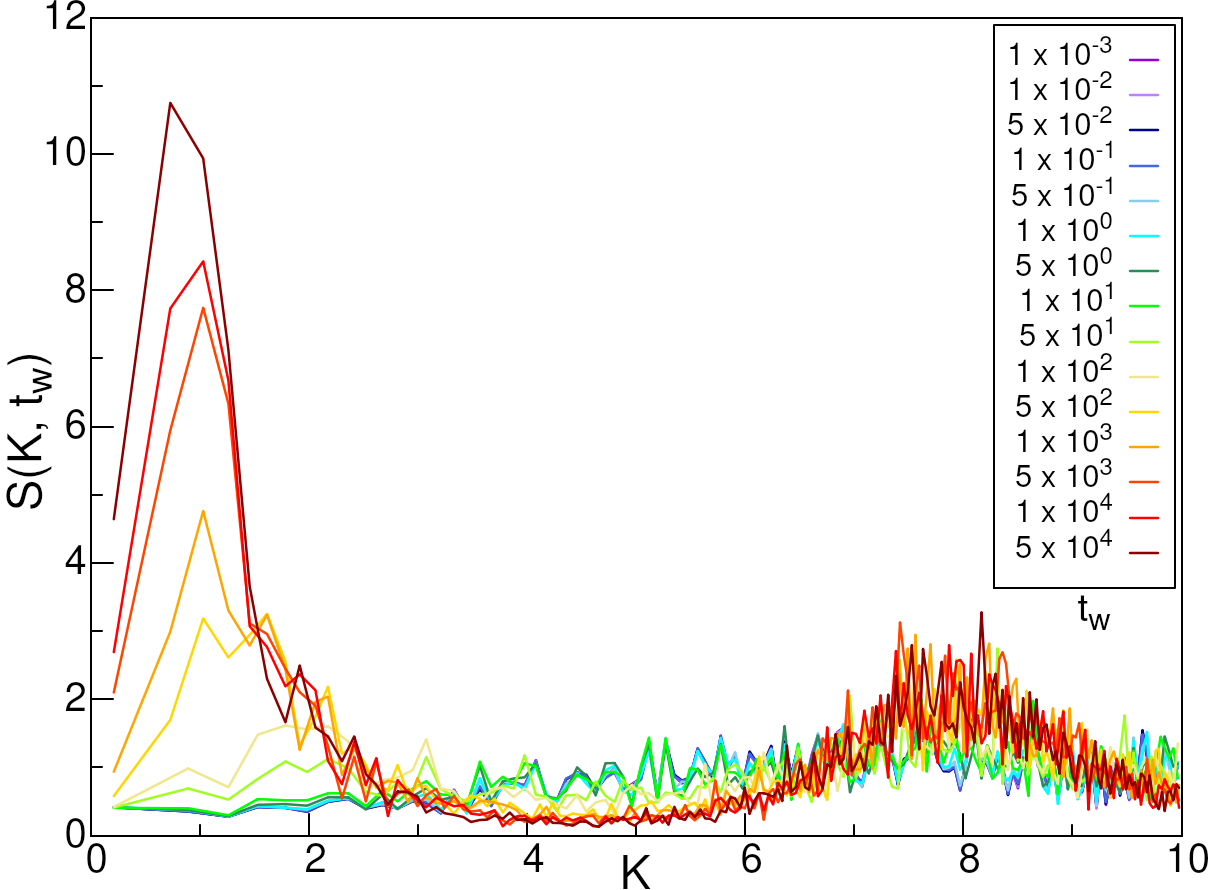}
    \caption{The evolution of the static structure factor with wave vector $K$ at different waiting times is shown for $T = 0.10$. With increasing $t_w$, the $1^\text{st}$ peak of $S(K, t_w)$ is shifts to lower $K$ values, that indicates the growth of local structural ordering.}
    \label{fig_appendix_b}
\end{figure}

\section*{Appendix B: Static Structure Factor}
The waiting time dependence of the static structure factor~\cite{patrick_jcp_2017, Patrick_Faraday-Discussions_2012} in reciprocal space for an out-of-equilibrium system of $N$-point particles is defined as,
\begin{equation}
    S(K, t_w) = \Bigg[\frac{1}{N} \sum_{j,k = 1}^N e^{-i\vec{K} \cdot \big(\vec{r}_k(t_w) - \vec{r}_j(t_w)\big)}\Bigg],
\end{equation}
where the term $\big(\vec{r}_k(t_w) - \vec{r}_j(t_w)\big)$ denotes the displacement vector of the $k^\text{th}$ particle relative to the $j^\text{th}$ particle measured at the waiting time $t_w$, the $\vec{K}$ is being the wave vector, and $[...]$ denotes an average over multiple realizations. In a finite simulation box of linear dimension $l$ with periodic boundary conditions, the allowed wave vectors are restricted to $\vec{K} = \frac{2\pi}{l}(n_x \hat{x} + n_y \hat{y} + n_z \hat{z})$, where $n_x$, $n_y$, and $n_z$ are non-zero integers. Thus, the smallest nonzero wave vector accessible in the simulation is $\Delta K = 2\pi/l$. To obtain an isotopically averaged static structure factor, we average over wave vectors with the same magnitude $|\vec{K}|$.

\vskip+0.04in
In Fig.~\ref{fig_appendix_b}, we show the waiting time evolution of $S(K,t_w)$ for different values of the wave vector magnitude $|\vec{K}|$. It is evident that, at smaller waiting times $t_w$, no apparent peaks are observed, reflecting a disordered fluid state. As the system ages or approaches the gelation time $t_\text{gel} \sim 470$, an apparent $1^\text{st}$ peak appears in the low-wave-vector regime. This indicates that, as the system approaches the corresponding gelation time (i.e., $t_w \ge t_\text{gel}$), a system-spanning percolating network forms rapidly. As the system ages further, the $1^\text{st}$ peak of $S(K, t_w)$ shifts toward lower $K$ values, indicating that the gelation process is accompanied by a complex structural reorganization of the local particle environment. In Fig.~\ref{fig_appendix_b}, the $1^\text{st}$ peak of $S(K, t_w)$ at the largest waiting time $t_w = 5 \times 10^4$, appears around the wave vector $K^* = 0.74$. At this characteristic wave vector $K^*$, the self-intermediate scattering function $F_s(K, t_w, t)$ (see definition in the Model and Methods section) does not progressively decay to zero with increasing waiting time $t_w$. Thus, we compute $F_s(K, t_w, t)$ at $K = 2.50$ and $10.00$, to extract the corresponding relaxation timescale $\tau(t_w)$. Conversely, Fig.~\ref{fig_appendix_b} shows an additional peak or the $2^\text{nd}$ peak, in the large wave-vector regime and the amplitude of this $2^\text{nd}$ peak increases with $t_w$. This enhancement of the $2^\text{nd}$ peak suggests the development of the medium-range structural correlations in the evolving percolated gel network~\cite{patrick_jcp_2017}.

\begin{figure*}
    \centering
    \includegraphics[page=1, trim=4.5cm 19cm 6cm 0cm, clip, width=1.00\textwidth]{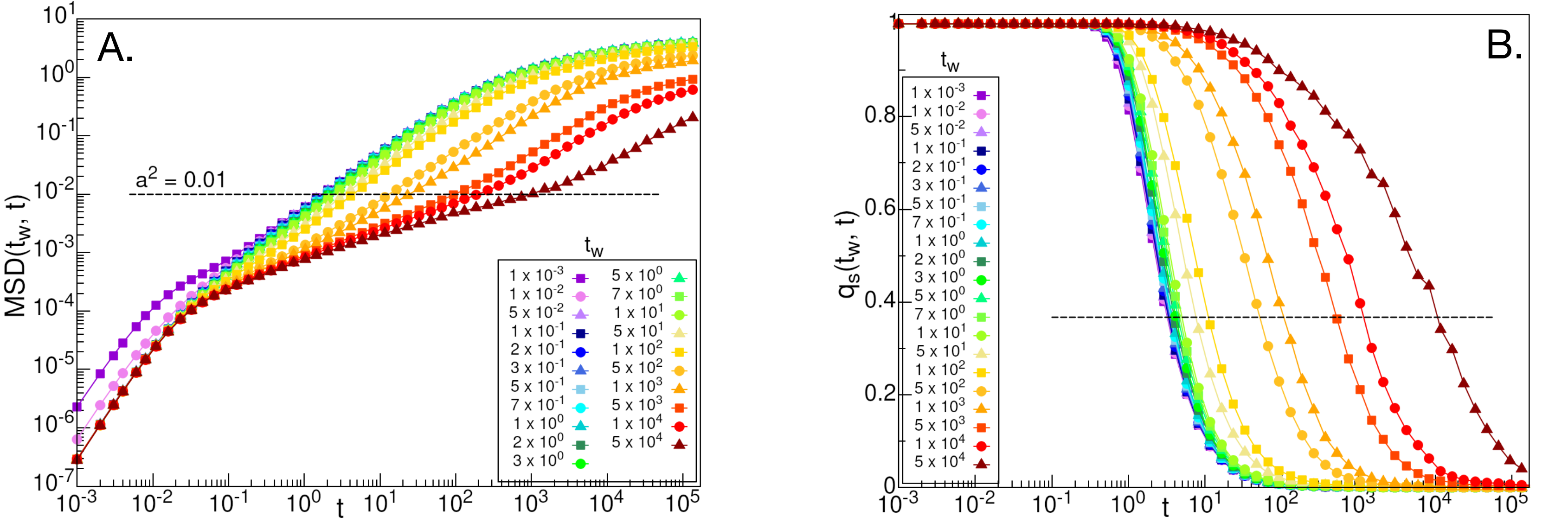}
    \caption{Temporal evolution of (A) the mean-squared displacement and (B) the overlap function for different waiting times at temperature $T = 0.10$. We determine the plateau value $a^2 = 0.01 \bar{\sigma}^2$ from the MSD at the largest waiting time. In (B), the black dotted line shows the $1/e$ value that is used to calculate the structural relaxation time $\tau(t_w)$ for the corresponding waiting time $t_w$.}
    \label{fig_appendix_c}
\end{figure*}

\section*{Appendix C: Mean Squared Displacement and Overlap Function}
The evolution of particle dynamics with the waiting time $t_w$ is characterized by calculating the mean-squared displacement (MSD)~\cite{olivier_prl_2023} and the self-overlap function~\cite{santu_arxiv_2025}, which for an $N$-point particle system undergoing the out-of-equilibrium dynamics, are defined as
\begin{equation}
    \text{MSD}(t_w, t) = \Bigg[\frac{1}{N} \sum_{i=1}^N \big|\vec{r}_i(t_w+t) - \vec{r}_i(t_w)\big|^2 \Bigg],
\end{equation}
and
\begin{equation}
    q_s(t_w, t) = \Bigg[\frac{1}{N} \sum_{i=1}^N w\big(\big|\vec{r}_i(t_w+t) - \vec{r}_i(t_w)\big|\big) \Bigg].
\end{equation}
Here, the term $\big|\vec{r}_i(t_w+t) - \vec{r}_i(t_w)\big|$ denotes the displacement of the $i^\text{th}$ particle within the time interval of $t_w+t$ and $t_w$. The window function is defined earlier as $w(x) = \Theta(a - x)$, with $\Theta$ being the Heaviside step function. The parameter $``a"$ is chosen to be comparable to the localization length associated with the plateau height of the mean-squared displacement (MSD) in the large $t_w$ regime. We introduce $a$ to eliminate decorrelation arising from the vibrational motion of particles within the cages formed by their nearest neighbours. Figures~\ref{fig_appendix_c}(A) and \ref{fig_appendix_c}(B) show the time evolution of the MSD and the overlap function, respectively, in an out-of-equilibrium dynamics. In Fig.~\ref{fig_appendix_c}(A), the plateau height is found to be $a^2 = 0.01\bar{\sigma}^2$ for the largest waiting time $t_w = 5 \times 10^4$. The relaxation timescale at different waiting time $t_w$ is defined as $q_s(t_w, t = \tau(t_w)) = 1/e$.

\begin{figure*}
    \centering
    \includegraphics[page=1, trim=0cm 21.5cm 0cm 0cm, clip, width=1.00\textwidth]{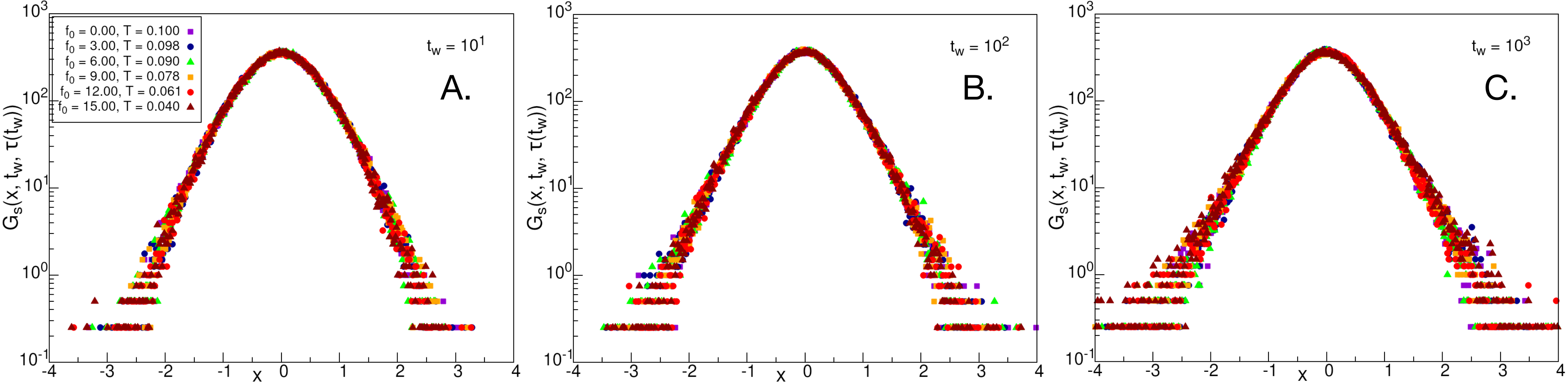}
    \caption{The spatial extent of the Van Hove function at $T_\text{eff} = 0.10$ for different values of $T$ and $f_0$, with $\tau_p=10^0$ and $\rho_a=0.10$ held fixed, is shown for waiting times (A) $10^1$, (B) $10^2$, and (C) $10^3$. The function $G_s(x, t_w, \tau(t_w))$ exhibits a crossover from Gaussian to non-Gaussian behavior as $t_w$ increases.}
    \label{fig_appendix_d}
\end{figure*}

\section*{Appendix D: Van Hove Function}
The non-stationary variant of the self-part of the Van Hove function $G_s(x,t_w,t)$, which characterizes the probability distribution of particle displacements over a time interval $t$, measured from the waiting time $t_w$, is defined as,
\begin{equation}
    G_s(x, t_w, t) = \Bigg[\sum_{i = 1}^N \delta\big(x - \left(x_i(t_w + t) - x_i(t_w)\right)\big)\Bigg],
\end{equation}
where $\left(x_i(t_w + t) - x_i(t_w)\right)$ denotes the displacement of the $i^\text{th}$ particle in the $x$-direction within the time interval of $t_w + t$ to $t_w$, and $\delta(\cdot)$ is the Dirac delta function. We choose the time interval $t$ as the structural relaxation time at the waiting time $t_w$. The self-part of the Van Hove function provides information about the distribution of individual particle displacements and can characterize the system's spatial heterogeneity. The results for different waiting times with the similar $T_\text{eff} = 0.10$ for different combinations of $T$ and $f_0$ are reported in Figs.\ref{fig_appendix_d}(A\textendash C). At each waiting time, the spatial extent of $G_s(x, t_w, \tau(t_w))$ remains similar across different choices of $T$ and $f_0$, indicating similar spatial heterogeneity in particle mobility.

\end{document}